\documentclass[extra,mreferee]{gji}
\usepackage{timet}
\usepackage{graphicx}
\usepackage{multirow}
\usepackage{multicol}
\usepackage{lipsum}
\usepackage{timet}
\usepackage{epsfig}
\usepackage{amsmath}
\usepackage{amsfonts}
\usepackage{amssymb}
\usepackage{color}
\usepackage[overload]{empheq}
\usepackage{float}
\usepackage{subcaption}
\usepackage{graphics}
\usepackage{xcolor,graphicx}
\usepackage{csquotes}
\usepackage{mathtools}
\usepackage{optidef}
\usepackage{hyperref}
\usepackage{algorithm}
\usepackage{algorithmic}
\usepackage{rotating}
\usepackage{multirow}
\usepackage{adjustbox}
\usepackage{lscape}
\usepackage{mathtools}
\hypersetup{colorlinks,linkcolor={blue},citecolor={blue},urlcolor={black}}
\usepackage{enumitem}
\usepackage{lipsum}
\usepackage{amsfonts}
\usepackage{graphicx}
\usepackage{mathrsfs}
\usepackage{epstopdf}
\usepackage{multirow} % Add this package
\usepackage{nicefrac}
\usepackage{algorithmic}
\usepackage{algorithm}
\usepackage{ragged2e}
\ifpdf
  \DeclareGraphicsExtensions{.eps,.pdf,.png,.jpg}
\else
  \DeclareGraphicsExtensions{.eps}
\fi
\usepackage{amsmath, amssymb}

\usepackage{tikz}
\usepackage{hyperref}
\usetikzlibrary{mindmap}
\usepackage{bm}
\newcommand{\bA}{\mathbf{A}}

\newcommand{\bb}{\mathbf{b}}

\newcommand{\bD}{\mathbf{D}}
\newcommand{\bd}{\mathbf{d}}

\newcommand{\bI}{\mathbf{I}}
\newcommand{\bL}{\mathbf{L}}
\newcommand{\bS}{\mathbf{S}}

\newcommand{\bH}{\mathbf{H}}

\newcommand{\m}{\mathbf{m}}
\newcommand{\bP}{\mathbf{P}}

\newcommand{\bu}{\mathbf{u}}

\newcommand{\br}{\mathbf{r}}

\newcommand{\sdual}{\boldsymbol{\lambda}}
\newcommand{\sdualp}{\boldsymbol{\nu}}
\newcommand{\Eps}{\bm{\varepsilon}}
\usepackage{subfiles}

\usepackage{booktabs}

\newcommand{\fref}[1]{Figure \ref{#1}}                  % Figure ref
\newcommand{\eref}[1]{\eqref{#1}}                  % equation ref

\DeclareMathOperator*{\minimize}{minimize}

\DeclareMathOperator*{\argmin}{argmin}

\DeclareMathOperator*{\argminimax}{argminimax}

\ifpdf
  \DeclareGraphicsExtensions{.eps,.pdf,.png,.jpg}
\else
  \DeclareGraphicsExtensions{.eps}
\fi

\usepackage{placeins} % Provides \FloatBarrier command
\usepackage{afterpage} % For \afterpage command
\usepackage{appendix}
\usepackage{amsopn}

\usepackage{titlesec}
\titleformat{\paragraph}
  {\normalfont\normalsize\bfseries}{\theparagraph}{1em}{}
\titlespacing{\paragraph}{0pt}{3.25ex plus 1ex minus .2ex}{1.5ex plus .2ex}

\usepackage{soul}
\usepackage{optidef}

\title[Penalty Parameter Selection for FWI]{Automatic Penalty Parameter Selection by Residual Whiteness Principle (RWP) and GCV for Full Waveform Inversion}

\usepackage{fancyhdr}
\makeatletter
\renewcommand\maketitle{%
  \thispagestyle{empty}%
  \begin{center}%
    {\huge\bfseries\@title\par}%
    \vspace{1.5em}%
    {\Large\@author\par}%
  \end{center}%
  \vspace{2em}%
}
\makeatother

\title[Penalty Parameter Selection for FWI]{Automatic Penalty Parameter Selection by Residual Whiteness Principle (RWP) and GCV for Full Waveform Inversion}

\author[Kamal Aghazade, Toktam Zand, Ali Gholami]{
  \begin{tabular}{c}
    \includegraphics[height=0.85em]{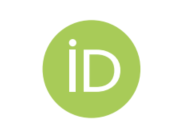}\ \href{https://orcid.org/0000-0002-2080-9697}{Kamal Aghazade} \\
    Institute of Geophysics, Polish Academy of Sciences, Warsaw, Poland \\
    \texttt{aghazade.kamal@igf.edu.pl} \\[0.5em]
    \includegraphics[height=0.85em]{orcid-iD_icon.png}\ \href{https://orcid.org/0000-0001-9834-0705}{Toktam Zand} \\
    Institute of Geophysics, Polish Academy of Sciences, Warsaw, Poland \\
    \texttt{tzand@igf.edu.pl} \\[0.5em]
    \includegraphics[height=0.85em]{orcid-iD_icon.png}\ \href{https://orcid.org/0000-0001-8403-1434}{Ali Gholami} \\
    Institute of Geophysics, Polish Academy of Sciences, Warsaw, Poland \\
    \texttt{agholami@igf.edu.pl}
  \end{tabular}
}

\begin{document}

\label{firstpage}
\maketitle
\thispagestyle{empty} 
% \keywords{Full waveform inversion; Wavefield reconstruction inversion; Extended-sources FWI; Augmented Lagrangian; Penalty parameter; Discrepancy principle; Generalized cross-validation; Residual whiteness principle.}

\begin{summary}
% \justifying
Full-waveform inversion (FWI) is a powerful seismic imaging technique used to estimate high-resolution physical properties of subsurface structures by minimizing the misfit between observed and modeled seismic data. FWI is inherently a highly non-linear and ill-posed inverse problem. Extended-source approaches, such as the augmented Lagrangian (AL) method, are employed to improve the convexity and robustness of the solution. A key component of this formulation is the penalty parameter ($\mu$), which controls the critical trade-off between fitting the observed data and satisfying the wave equation constraint, significantly influencing convergence, especially in the presence of noise. The practical challenge lies in selecting $\mu$. Traditional strategies like the Discrepancy Principle (DP) require an accurate estimate of the noise level ($\sigma$), which is often unknown or poorly characterized. Furthermore, trial-and-error tuning requires repeatedly solving the inverse problem, making it computationally prohibitive. To address these limitations and create a parameter-free, computationally efficient extended-source FWI algorithm, we propose integrating two data-driven parameter selection strategies—the Residual Whiteness Principle (RWP) and a stable variant of Generalized Cross-Validation (RGCV)—within the highly efficient multiplier-oriented formulation of AL. Specifically, we utilize a dual-space AL method. This framework is essential because it allows the background wave equation operator to remain fixed, requiring only a single LU matrix factorization per frequency inversion, which dramatically increases computational efficiency. This design enables the dynamic and efficient adjustment of $\mu$ within each iteration at negligible computational cost, making the resulting algorithm scalable and practical for large-scale applications.
Extensive numerical experiments on both acoustic and elastic FWI with white and colored noise show that combined with the dual-space formulation, RWP demonstrates excellent noise robustness and computational efficiency, making it a robust, automated solution for large-scale seismic inversion.
\end{summary}

\begin{keywords}
Full waveform inversion; Wavefield reconstruction inversion; Extended-sources FWI; Augmented Lagrangian; Penalty parameter; Discrepancy principle; Generalized cross-validation; Residual whiteness principle. 
\end{keywords}

\graphicspath{{./figures/}}

\section{INTRODUCTION}
Full-waveform inversion (FWI) is a powerful seismic imaging technique used to reconstruct subsurface physical properties \citep{Tarantola_1984_ISR, Pratt_1998_GNF, Operto_2023_FWI,Metivier_2025_HRM}, with broad applications in wave-based imaging \citep[e.g.,][]{Thrastarson_2022_DAG, Fachtony_2025_CO2}. FWI operates by iteratively minimizing the discrepancy (misfit) between simulated and observed seismic data. However, FWI is generally viewed as a highly non-linear and ill-posed inverse problem \citep[e.g.,][]{Mulder_2008_ESI,Kirsch_2014_STL}, necessitating the use of regularization techniques to stabilize the inversion and obtain meaningful solutions. Extended-source approaches, such as the augmented Lagrangian (AL) method \citep{Bertsekas_1982_COA}, offer a robust means of improving the problem's convexity \citep{Gholami_2024_FWI}. A key component of this formulation is the penalty parameter ($\mu$), which governs the critical trade-off between minimizing the data misfit (residual) and satisfying the wave equation constraint. The specific choice of this parameter significantly influences convergence and practical performance, particularly in the presence of noise
 \citep{VanLeeuwen_2013_MLM,Huang_2018_VSE, Aghamiry_2019_IWR, Gholami_2022_EFW,Lin_2023_FWR}.

The challenge lies in efficiently and reliably selecting the optimal $\mu$. Traditionally, $\mu$ is determined empirically, often set as a fraction of the maximum eigenvalue of the augmented wave-equation operator \citep[e.g.,][]{vanLeeuwen_2016_PMP, Aghamiry_2019_IWR, Operto_2023_FWI}. Alternatively, the Discrepancy Principle (DP) is used, which selects $\mu$ such that the norm of the data residual matches the estimated noise level \citep{Fu_2017_DPM, Gholami_2024_FWI, Symes_2025_NIM}. The DP approach is severely limited because it relies on an accurate estimate of the noise standard deviation ($\sigma$), which is often unknown or poorly characterized in practical applications. Furthermore, the DP considers only the zero-lag value of the residual autocorrelation, thereby neglecting potentially informative correlations at nonzero lags. Consequently, the conventional trial-and-error tuning or methods reliant on accurate $\sigma$ estimates become computationally prohibitive. Therefore, developing efficient, data-driven strategies for penalty parameter estimation is essential for computationally demanding FWI problems.

To address these limitations, we propose employing the Residual Whiteness Principle (RWP) \citep{Almeida_2013_PEB,Lanza_2013_VID} and the Robust Generalized Cross-Validation (RGCV) \citep{Lukas_2006_RGCV}. The primary contribution of this work is the integration of these two data-driven parameter selection strategies within the highly efficient multiplier-oriented formulation of the AL method \citep{Gholami_2024_FWI}, specifically utilizing the dual-space AL (Dual-AL) method \citep{Aghazade_2025_FAF}. This choice of framework is essential due to its superior computational efficiency. The Dual-AL method is rooted in the perspective that the Lagrange multipliers are the fundamental unknowns, allowing the background wave equation operator, $\bold{A}(\bold{m}^0)$, to remain fixed for each frequency inversion. This design only requires a single LU matrix factorization per frequency inversion. In contrast, standard primal algorithms necessitate repeated factorizations at every iteration. This crucial efficiency gain enables the dynamic and efficient update of the penalty parameter $\mu$ within each iteration at negligible computational cost, thereby making the resulting algorithm scalable and practical for large-scale applications.

RWP is an automatic, parameter-free strategy that overcomes DP's limitations by evaluating the full autocorrelation function \citep{Pragliola_2023_ADMM}. It selects the optimal $\mu$ by minimizing a measure of non-whiteness of the residual data. This measure is based on kurtosis minimization, forcing the residual to resemble white noise, which has a maximally sparse autocorrelation function. We also adopt RGCV, a stable extension of GCV, which addresses the instability of standard GCV for smaller datasets that often leads to under-regularization. Extensive numerical experiments on both acoustic and elastic FWI demonstrate that the combination of RWP/RGCV with the Dual-AL formulation exhibits exceptional noise robustness and computational efficiency. Specifically, RWP consistently achieves the closest approximation to the optimal parameter choice among the tested methods, making it a robust, automated solution for large-scale seismic inversion.

\section{Preliminaries}
This section establishes the mathematical context for solving ill-posed inverse problems and reviews the specific FWI formulations and parameter selection principles employed in this work.
\subsection{The Regularized Inverse Problem}
We consider the problem of estimating an unknown model $\mathbf{m}$ from noisy observations $\mathbf{d}$, typically formulated as a discrete ill-posed inverse problem. FWI is a nonlinear and ill-posed problem that requires regularization to stabilize the solution. Stabilization is commonly introduced within the framework of a generalized Tikhonov formulation:
\begin{equation} \label{Gm_obj}
\minimize_{\mathbf{m}}  \frac{1}{2} \|\mathbf{G}(\mathbf{m}) - \mathbf{d}\|^2 + \frac{\mu}{2} \mathcal{R}(\mathbf{m}) 
\end{equation}
Here, $\mathbf{G}$ is the forward operator, $\mathcal{R}(\mathbf{m})$ is the regularization functional that encodes prior structural information about the model (e.g., promoting smooth or piecewise-constant features), and $\mu > 0$ is the regularization parameter controlling the trade-off between data fidelity and regularization strength. Common regularization techniques include Tikhonov regularization (which promotes smoothness) \citep{Tikhonov_1977_SIP}, Total Variation (TV) regularization (which preserves sharp edges) \citep{Rudin_1992_NTV}, or their combination, Tikhonov-TV (TT) regularization (which preserves both the smooth parts and sharp edges \citep{Gholami_2013_BCT}.

\subsection{Automatic Parameter Selection Methods}
The central difficulty in regularization is the selection of the optimal parameter $\mu$, which dictates the trade-off between stability and fidelity. We review the criteria utilized for automatic parameter selection: the Discrepancy Principle (DP), Generalized Cross-Validation (GCV), and Residual Whiteness Principle (RWP).

The DP is a widely used deterministic method that selects the parameter $\mu$ such that the norm of the data residual ($\|\mathbf{r}_{\mu}\|_2$) matches the estimated noise level \citep{Morozov_1966_SFE,Bredies_2013_DPA,Jahn_2021_MDP}. A key technical limitation of the DP is that it relies on an accurate estimate of the noise standard deviation ($\sigma^2$), which is often unknown. Furthermore, it considers only the zero-lag value of the residual autocorrelation, potentially neglecting informative correlations at nonzero lags. GCV is a data-driven method that minimizes an objective function related to the prediction error and does not require knowledge of the noise variance ($\sigma^2$) \citep[e.g., ][]{Wahba_1977_PAS,Lukas_1998_CPC,Haber_2000_GCV, Gholami_2013_FBD}. However, standard GCV can be unreliable for small or moderate-sized datasets, often leading to an overly small regularization parameter $\mu$. To mitigate this instability, the Robust GCV (RGCV) extension is adopted \citep{Lukas_2006_RGCV}. The RWP is a recently developed, automatic, parameter-free strategy proposed when the data noise is assumed to be additive white Gaussian noise (AWGN). The RWP assumes that the predicted residual ($\mathbf{r}_{\mu}=\mathbf{G}(\mathbf{m}_{\mu}) - \mathbf{d}$) of an optimally regularized solution should resemble a realization of white noise, implying that its autocorrelation function should be maximally sparse (a spike), as measured by the kurtosis \citep{Almeida_2013_PEB,Lanza_2013_VID}.

\subsection{Full Waveform Inversion Formulations}

\subsubsection{Forward Problem.}
We present algorithms for both acoustic-and elastic FWI. While the formulations are derived in the frequency domain, the methodology and results are readily extendable to the time domain.

For acoustic media, the frequency-domain wave propagation is described by the Helmholtz equation:
\begin{equation}\label{AC_forward}
    (\omega^2 \text{diag}(\m)+\Delta)\bu_{s} = \bb_{s},
\end{equation}
where $\m \in \mathbb{R}^{n}$ denotes the model parameters (defined as the reciprocal of squared velocity), $\bu_s \in \mathbb{C}^{n},~s=1,\dots,n_s,$ are the wavefields corresponding to the source terms $\bb_s \in \mathbb{C}^{n}$ ($n_s$ is the number of sources), $\text{diag}(\bullet)$ is the diagonal operator, and $\Delta$ represents the Laplace operator. Each wavefield $\bu_s$ satisfies the discretized wave equation
$
\bA(\m)\bu_s=\bb_s,
$
where $\bA(\m)=(\omega^2 \text{diag}(\m)+\Delta) \in \mathbb{C}^{n \times n}$ serves as the wave propagation operator for the angular frequency $\omega$.

For elastic media, we consider the isotropic frequency-domain formulation for 2D medium:
\begin{equation}\label{El_forward}
\begin{aligned}
&\bm{\rho} \omega^{2}\bu_{x,s}\!\!+\!(\Tilde{\bm{\lambda}}\!+\!2\Tilde{\bm{\mu}})\partial_{xx} \bu_{x,s}\!\!+\!\Tilde{\bm{\mu}} \partial_{zz} \bu_{x,s} \! +\! (\Tilde{\bm{\lambda}}+\Tilde{\bm{\mu}})\partial_{xz}\bu_{z,s} \!= \!\bb_{x,s}, \\
&(\Tilde{\bm{\lambda}}+\Tilde{\bm{\mu}})\partial_{xz} \bu_{x,s} + \bm{\rho} \omega^{2} \bu_{z,s}\!\!+\!(\Tilde{\bm{\lambda}}\!+\!2\Tilde{\bm{\mu}})\partial_{zz} \bu_{z,s}\!\!+\!\Tilde{\bm{\mu}} \partial_{xx}\bu_{z,s} \!=\! \bb_{z,s},
\end{aligned}
\end{equation}
where $\bm{\rho} \in \mathbb{R}^{n}$ is the density, $\Tilde{\bm{\lambda}} \in \mathbb{R}^{n}$ and $\Tilde{\bm{\mu}} \in \mathbb{R}^{n}$ are the Lamé parameters, $\bu_{x,s} \in \mathbb{C}^{n}$ and $\bu_{z,s} \in \mathbb{C}^{n}$ denote horizontal and vertical displacements, and $\bb_{x,s} \in \mathbb{C}^{n}, \bb_{z,s} \in \mathbb{C}^{n}$ are the source terms. In addition, the notation $\partial_{ij}=\frac{\partial^2}{\partial_i \partial_j}$ represents the second-order partial derivative with respect to $i$ and $j$.
In this work, we adopt the constant-density assumption ($\bm{\rho}=1$). Using the definitions of P-wave velocity $\text{V}_\text{P} = \big( (\Tilde{\bm{\lambda}}+2\Tilde{\bm{\mu}})/\bm{\rho}\big)^{1/2}$ and S-wave velocity $\text{V}_\text{S} = (\Tilde{\bm{\mu}}/\bm{\rho})^{1/2}$, the model parameters are expressed as
$
\m = (\m_\text{P} := \text{V}_\text{P}^2,~\m_\text{S} := \text{V}_\text{S}^2).
$
Accordingly, the elastic system can be rewritten in matrix-vector form as
\begin{equation}
    \bA(\m_\text{P},~\m_\text{S})
  \begin{bmatrix}
\bu_{x,s}\\
\bu_{z,s}
\end{bmatrix} 	
= 
\begin{bmatrix}
\bb_{x,s}\\
\bb_{z,s}
\end{bmatrix}, \quad s=1,\dots, n_s,
\end{equation}
which can also be expressed in compact notation as
$
\bA(\m)\bu_s=\bb_s,
$
with $\m=(\m_\text{P},~\m_\text{S}) \in \mathbb{R}^{2n}$ , $\bu_s=(\bu_{x,s},~\bu_{z,s}) \in \mathbb{C}^{2n}$, and $\bb_s=(\bb_{x,s},~\bb_{z,s}) \in \mathbb{C}^{2n}$.  
Both acoustic and elastic wave-equation operators are constructed with sufficient numerical accuracy and equipped with suitable boundary conditions. For details on the discretization procedure, we refer the reader to \citep{Chen_2013_OFD,Chen_2016_MFE}.

\subsubsection{Inverse Problem.}
In the frequency domain, FWI can be formulated as the following nonlinearly constrained optimization problem \citep{Haber_2000_OTS}
\begin{equation}\label{main}
    \min_{\m,\bu_s}~\frac12\sum_{s=1}^{n_s}\|\bP_s\bu_s-\bd_s\|_2^2~~\text{s.t.}~~ \bA(\m)\bu_s=\bb_s,~s=1,...,n_s
\end{equation}
where $\|\cdot\|_2$ denotes the vector $2$-norm, $\bd_s$ are the observed data, and $\bP_s$ are the sampling operator. 
Different strategies exist to handle the complexity and size of the wave propagation constraint, as especial cases of the following proximal Lagrangian function \citep{Gholami_2024_FWI}
\begin{equation}\label{AL_proximal}
 \min_{\m,\bu_s}\max_{\sdualp_s} \hat{\mathcal{L}}(\m,\bu_s,\sdualp_s,\hat{\sdualp}_s) = \sum_{s=1}^{n_s} \mathcal{L}(\m,\bu_s,\sdualp_s)-\sum_{s=1}^{n_s}\frac{1}{2\mu_s} \|\sdualp_s - \hat{\sdualp}_s\|_2^2
\end{equation}
where $\mathcal{L}$ is the standard Lagrangian function \citep{Haber_2000_OTS}
\begin{equation}
    \mathcal{L}(\m,\bu_s,\sdualp_s)=\frac{1}{2}\|\bP_s\bu_s-\bd_s\|_2^2+\langle\sdualp_s,\bA(\m)\bu_s-\bb_s\rangle.
\end{equation}
$\sdualp_s$ are the Lagrange multiplier (or source multiplier) vectors associated with the wave-equation constraints, and $\langle \cdot, \cdot \rangle$ denotes the inner product. The second term in equation \eqref{AL_proximal} introduces a source-space regularization that penalizes deviations of the Lagrange multiplier $\sdualp_s$ from a reference estimate $\hat{\sdualp}_s$, which varies with iterations and can be estimated as \citep{Powell_1969_NLC}:
\begin{equation}
    \hat{\sdualp}_s^{k+1}=\hat{\sdualp}_s^{k}+\mu_s^k (\bA(\m^{k+1})\bu_s^{k+1}-\bb_s),\quad s=1,...,n_s.
\end{equation}
Traditional FWI formulations can be recovered as special cases of \eqref{AL_proximal}. 
Setting $\mu_s\to \infty$ leads to a reduced objective function that depends only on the model parameters $\m$ \citep[e.g.,][]{Tarantola_1984_ISR, Pratt_1998_GNF}. While effective, this reduced formulation often results in an ill-conditioned problem and complicates step-length tuning.
Also, by setting $\hat{\sdualp}_s = 0$, the proximal Lagrangian \eqref{AL_proximal} reduces to classical quadratic-penalty formulations \citep[e.g.,][]{Abubakar_2009_FDC, VanLeeuwen_2013_MLM, vanLeeuwen_2016_PMP}, where the wave-equation constraint is enforced approximately through the penalty term.

%\item  Augmented Lagrangian Method: This is the specific framework adopted for computational efficiency. In this dual domain approach, the model parameters ($\m$) and the wavefield ($\bu$) are expressed as implicit functions of the Lagrange multipliers ($\hat{\sdualp}$). This dependence allows for the elimination of $\m$ and $\bu$ as optimization variables, leading to a dual AL function dependent solely on $\hat{\sdualp}$. Critically, this strategy allows the forward and adjoint wavefields to be computed using a fixed background model ($\m_0$). The fixed wave equation operator enables the reuse of a single LU matrix factorization per frequency inversion, providing superior computational efficiency \citep{Aghazade_2025_FAF}.
%\end{itemize}

% where $\|\cdot\|_2$ denotes the vector $2$-norm, $\bd_s$ are the observed data, and $\bP_s$ are the sampling operator. 
% The optimization problem in \eref{main} can be addressed using various strategies, including the reduced-space approach \citep{Tarantola_1984_ISR,Pratt_1998_GNF}, penalty method \citep{Abubakar_2009_FDC,VanLeeuwen_2013_MLM}, and the multiplier or augmented Lagrangian (AL) method \citep{Aghamiry_2019_IWR,Gholami_2024_FWI}, each representing well-established frameworks. In this paper, we do not revisit the detailed derivations of these algorithms, as they are thoroughly presented in \citep{Gholami_2024_FWI}. Instead, we adopt the multiplier-oriented formulation of the AL, which offers a unified algorithmic framework encompassing all three approaches and is applicable in both time and frequency domains.

%
\section{Theory}

The  ADMM iteration for solving the optimization problem \eqref{main}  can be written as \citep{Gholami_2024_FWI}
\begin{subequations}\label{ALM}
\begin{align} 
(\bu_s^{k+1},\sdualp_s^{k+1})&=\argminimax_{\bu,\sdualp}\sum_{s=1}^{n_s} \mathcal{L}(\m^{k},\bu_s,\sdualp_s)-\sum_{s=1}^{n_s} \frac{1}{2\mu_s^k}\|\sdualp_s-\hat{\sdualp}_s^{k}\|_2^2 \label{ALM_saddle}, \\
\m^{k+1}&=\m^{k}- \left(\sum_{s=1}^{n_s} \bL({\bu^{k+1}_s})^\top \bL({\bu^{k+1}_s})\right)^{-1}\left(\sum_{s=1}^{n_s} \frac{1}{\mu_s^k}\bL({\bu^{k+1}_s})^{\top} \sdualp^{k+1}_s\right),  \label{ALM_model}\\
\hat{\sdualp}_s^{k+1}&=\hat{\sdualp}_s^{k}+\mu_s^k (\bA(\m^{k+1})\bu_s^{k+1}-\bb_s),\quad s=1,...,n_s, \label{ALM_dual}
\end{align}
\end{subequations}
where $\bL({\bu^{k+1}_s}) = \frac{\partial \bA(\m)}{\partial \m}\bu_s^{k+1}$. %For updating the model perturbation in \eqref{ALM_model}, the gradient is computed using the adjoint-state method \citep{Plessix_2006_RAS}, employing the pseudo-Hessian approximation  ($\bL(\bu)^{\top}\bL(\bu)$) \citep{Shin_2001_IAP}.
The parameter $\mu_s^k > 0$ acts as the regularization or penalty weights, controlling the strength of this enforcement and contributing to the stabilization of the Lagrangian function. Unlike \cite{Gholami_2024_FWI}, who use a fixed penalty parameter, this paper introduces a dynamic adjustment strategy, allowing $\mu_s^k$ to vary for each source and at each iteration. This adaptive scheme provides greater flexibility and improves the inversion’s robustness and convergence. The two most challenging steps in the algorithm described by equation \eqref{ALM} are: (1) solving the saddle point problem in \eqref{ALM_saddle}, and (2) selecting appropriate values for the regularization parameters $\mu_s^k$. We first address efficient strategies for solving \eqref{ALM_saddle}, while the selection of regularization parameters is discussed in detail in Section \ref{Sec_PPS}.

Computing the gradient of the objective function in \eref{ALM_saddle} with respect to $\bu_s$ and $\sdualp_s$, and setting it to zero, leads to the following saddle-point system:

\begin{equation}
    \begin{pmatrix}
    \bP_s^{\top}\bP_s & \bA(\m^k)^{\top}\\
    \bA(\m^k) & \mu_s^k \bI
    \end{pmatrix}
    \begin{pmatrix}
    \bu_s^{k+1}\\
    \sdualp_s^{k+1}
    \end{pmatrix}=
    \begin{pmatrix}
    \bP_s^{\top}\bd_s\\
    \bb_s- \frac{1}{\mu_s^k}\hat{\sdualp}_s^k
    \end{pmatrix}.
\end{equation}
Two different strategies—the wavefield-oriented approach and the multiplier-oriented approach—can be used to solve this system, each leading to distinct but mathematically equivalent iterative procedures (see \cite{Gholami_2024_FWI}, their Section 3.3). In this work, we adopt the multiplier-oriented approach and introduce the scaled variables: $\sdual_s^k = \frac{1}{\mu_s^k}\sdualp_s^k, \sdual_s^{k+1} = \frac{1}{\mu_s^k}\sdualp_s^{k+1}, \Eps_s^k = \frac{1}{\mu_s^k}\hat{\sdualp}_s^k$ and $\Eps_s^{k+1} = \frac{1}{\mu_s^k}\hat{\sdualp}_s^{k+1}$. The resulting iteration is as follows:
\begin{subequations}
\label{AL_alg}
\begin{align} 
\sdual_s^{k+1}&=(\bS_{s}^k)^{\top} \left( \bS_{s}^k(\bS_{s}^k)^{\top}+\mu_s^k \bI \right) ^{-1} (\bd_s-\bS_{s}^k[\bb_s-\Eps_s^{k}]), \quad s=1,...,n_s\label{ALM1nu} \\ 
\bu_s^{k+1}&=\bA(\m^k)^{-1}(\bb_s+\sdual_s^{k+1}- \Eps_s^{k}), \quad s=1,...,n_s\label{ALM1u} \\
\m^{k+1}&=\m^{k}- \left(\sum_{s=1}^{n_s} \bL({\bu^{k+1}_s})^\top \bL({\bu^{k+1}_s})\right)^{-1} \left(\sum_{s=1}^{n_s} \bL({\bu^{k+1}_s})^{\top} \sdual_s^{k+1}\right), \label{ALM22}\\
\Eps_s^{k+1}&=\Eps_s^{k}+\bA(\m^{k+1})\bu_s^{k+1}-\bb_s,\quad s=1,...,n_s \label{ALM33}
\end{align}
\end{subequations}
where $\bS_{s}^{k}=\bP_s\bA(\m^{k})^{-1}$. 

\subsection{An alternative dual-space algorithm}
The main computational bottleneck of the algorithm presented in \eref{AL_alg} lies in the computation of the forward and adjoint wavefields, $\bu_s$ and $\sdual_s$, both of which depend on the model parameters that are updated at each iteration. This dependency requires repeated LU factorizations of the wave equation operator, which is computationally expensive. A practical remedy for this issue is to use the dual-AL approach proposed in \citep{Aghazade_2025_FAF}. In this dual approach, the forward and adjoint wavefields are computed using a fixed background model $\m^0$, allowing for a single precomputed LU factorization to be reused throughout the iterations. Meanwhile, the Lagrange multipliers are still computed using the updated parameters, as specified in the following iteration:
% \vspace{0.4cm}
% \begin{subequations}\label{DALM}
% \begin{align} 
% (\bu_s^{k+1},\sdualp_s^{k+1})&=\argminimax_{\bu,\sdualp}\sum_{s=1}^{n_s} \mathcal{L}(\m^0,\bu_s,\sdualp_s)-\sum_{s=1}^{n_s} \frac{1}{2\mu_s^k}\|\sdualp_s-\hat{\sdualp}_s^{k}\|_2^2, \label{DALM1} \\
% \delta \m^{k+1}&=- \dfrac{\sum_{s=1}^{n_s}\langle \bL({\bu^{k+1}_s}), \frac{1}{\mu_s^k} \sdualp^{k+1}_s\rangle }{\sum_{s=1}^{n_s} \bL({\bu^{k+1}_s})^\top \bL({\bu^{k+1}_s})}, \label{ALM2}\\
% \hat{\sdualp}_s^{k+1}&=\hat{\sdualp}_s^{k}+\mu_s^k (\bA(\m^0+\delta \m^{k+1})\bu_s^{k+1}-\bb_s),\quad s=1,...,n_s \label{ALM3}.
% \end{align}
% \end{subequations}
% Solving the saddle point subproblem in \eref{DALM1} leads to
\begin{subequations}
\label{DAL_alg}
\begin{align} 
\sdual_s^{k+1}&=(\bS_{s}^0)^{\top}(\bS_{s}^0(\bS_{s}^0)^{\top}+\mu_s^k \bI)^{-1} (\bd_s-\bS_{s}^0[\bb_s-\Eps_s^{k}]), \quad s=1,...,n_s\label{DALM1nu} \\
\bu_s^{k+1}&=(\bA^0)^{-1}(\bb_s+\sdual_s^{k+1}- \Eps_s^{k}), \quad s=1,...,n_s\label{DALM1u} \\
\delta \m^{k+1}&=-\left(\sum_{s=1}^{n_s} \bL({\bu^{k+1}_s})^\top \bL({\bu^{k+1}_s})\right)^{-1}\left(\sum_{s=1}^{n_s} \bL({\bu^{k+1}_s})^{\top} \sdual_s^{k+1}\right),  \label{DALM22}\\
\Eps_s^{k+1}&=\Eps_s^{k}+\bA(\m^0+\delta \m^{k+1})\bu_s^{k+1}-\bb_s,\quad s=1,...,n_s, \label{DALM33}
\end{align}
\end{subequations}
where $\bA^0\equiv \bA(\m^0)$ and $\bS_{s}^0=\bP_s(\bA_s^0)^{-1}$.
\vspace{0.5cm}
Algorithms~\ref{alg1} and \ref{alg2} summarize the standard and dual-AL formulations, respectively. Algorithm~\ref{alg1} requires repeated LU factorizations of $\bA(\m^k)$ as the model updates, iterating over frequencies and alternately updating the backpropagating wavefields, $\sdual_s^{k+1}$, forward wavefields, $\bu_s^{k+1}$, model parameters, $\m^{k+1}$, and Lagrange multipliers, $\Eps_s^{k+1}$. In contrast, Algorithm~\ref{alg2} achieves efficiency by fixing the background model $\m^0$, requiring only one LU factorization of $\bA(\m^0)$ per frequency and updating model perturbation, $\delta\m^{k+1}$, instead of direct model parameters, while still maintaining the constraint enforcement through the updated Lagrange multipliers that account for the current model estimate $\m^0 + \delta\m^{k+1}$. 

The first subproblem in both algorithms, \eref{ALM1nu} and \eref{DALM1nu}, is the key step that addresses the uncertainty in the observed data. As seen from these subproblems, the penalty parameter $\mu_s^k$ plays a crucial role in ensuring stable computation of backpropagating wavefield $\sdual_s^{k+1}$ while allowing the data to be fitted at the noise level. An appropriate choice of $\mu_s^k$ at this stage enables effective suppression of noise, which would otherwise degrade the performance of the subsequent subproblems. In both algorithms, we adaptively adjust the penalty parameter $\mu_s^k$ using one of three strategies: DP, GCV, or RWP (described in the next section).

From a computational perspective, the dual formulation in Algorithm~\ref{alg2} provides significant efficiency gains. The costly LU decomposition of the wave operator, which scales as $\mathcal{O}(n^3)$, and the computation of the data-space Hessian ($\bS_s\bS_s^{\top}$) are performed only once per frequency. Subsequent wavefield updates reduce to efficient forward/backward substitutions, giving a total complexity of $\mathcal{O}(n_\omega \times n^3)$. In contrast, the standard augmented Lagrangian method requires repeated factorizations at every inner iteration, with complexity $\mathcal{O}(\texttt{maxit} \times n_\omega \times n^3)$, yielding a speedup roughly equal to the number of inner iterations per frequency. This makes the dual algorithm particularly advantageous for large-scale inversions.

%-------------------------------------------------------------------------------------------------------------------------------------
%                                                                   Penalty Parameter Selection (DP vs WP)
%-------------------------------------------------------------------------------------------------------------------------------------
\section{Regularization Parameter Selection} \label{Sec_PPS}
In this section, we describe three approaches for selecting the penalty parameter: the standard DP, GCV and RWP.
However, we will assume that the dependency of the variables on the source index $s$ and iteration number $k$ has been removed for the sake of simplicity and compact notation.
Before delving into the details of each method, it is important to note that the backpropagating wavefield 
${\sdual}$, defined in \eref{ALM1nu} or \eref{DALM1nu}, represents a least-squares solution to the following under-determined linear system:
\begin{equation}\label{Slambda}
    \bS\sdual_{\mu}=\delta \bd,
\end{equation}
where the right-hand side is the predicted data residual given by $\delta \bd={\bd}- {\bS}[\bb- {\Eps}]$. The parameter $\mu$ plays the role of a regularization parameter in solving \eqref{Slambda}. Specifically, as $\mu \to 0$ the solution ${\sdual}_{\mu}$ approaches the minimum-norm solution that satisfies \eqref{Slambda}. Conversely, as $\mu \to \infty$, the solution ${\sdual}_{\mu}$ tends to zero. The parameter $\mu$ must be carefully tuned to balance the trade-off between fitting the observed data and satisfying the wave equation. In the presence of noise, this tuning becomes even more critical to ensure that the data residual 
\begin{equation} \label{e}
    \bold{r}_{\mu} =\bS{\sdual}_{\mu}-\delta \bd,
\end{equation}
accurately models the noise component in the data and does not lead to overfitting or underfitting. Plugging for ${\sdual}_{\mu}$ from \eqref{ALM1nu} (or \ref{DALM1nu}) into \eqref{e} gives
\begin{align}
    \bold{r}_{\mu} &= \bS{\bS}^{\top} \left({\bS} {\bS}^{\top}+\mu{\bI}\right)^{-1}\delta \bd-\delta \bd \nonumber\\
     &= \left(\bS{\bS}^{\top} \left({\bS} {\bS}^{\top}+\mu{\bI}\right)^{-1}-\bI\right)\delta \bd \nonumber\\
     &= -\left(\frac{1}{\mu}{\bS} {\bS}^{\top}+{\bI}\right)^{-1}\delta \bd \label{e2}.
\end{align}
Using the eigenvalue decomposition of matrix ${\bS}{\bS}^{\top}$ in \eqref{e2},
\begin{equation} \label{eig}
{\bS}{\bS}^{\top}=\bold{V}\boldsymbol{\Sigma} \bold{V}^{\top},
\end{equation}
 where $\boldsymbol{\Sigma}$  is a diagonal matrix containing the eigenvalues, $\sigma_i, i=1,...,n_r$, each corresponding to the eigenvector in the same column $\bold{v}_i$ of the orthonormal matrix $\bold{V}$, the residual can be represented as
 \begin{align}
    \bold{r}_{\mu} = \sum_{i=1}^{n_r} -\frac{\bold{v}_i^{\top}\delta \bold{d}}{\sigma_i/\mu+1} \bold{v}_i .
\end{align}
\subsection{The method of discrepancy principle}
The DP principle \citep{Morozov_1966_SFE} selects the regularization parameter $\mu$ such that the norm of $\bold{r}_{\mu}$ matches the norm of the noise, $\eta$, affecting the data:
\begin{equation} \label{mu_dp}
 \mu =  \arg\min_{\mu} \phi^{\text{dp}}(\mu)=\frac12\left(\|\bold{r}_{\mu}\|_2^2-\eta^2\right)^2.
\end{equation}
In the case of Gaussian white noise, $\eta^2$ can be estimated as $\tau n_r \xi^2$ where $\xi^2$ is the noise variance and $\tau>1$ ($\tau=1$ if $\xi^2$ is accurately known). Therefore, the DP requires an accurate estimate of the noise level or standard deviation which is a major limitation \citep{Fu_2017_DPM}.
 % the objective function $\phi^{\text{dp}}$ in \eqref{mu_dp} can be written as
 % \begin{equation} \label{phi_dp}
 %     \phi^{\text{dp}}(\mu)=\frac12\left(\|W(\mu)\tilde{\delta d}\|_2^2-\eta^2\right)^2,
 %   % \phi^{\text{dp}}(\mu)=\frac12\left(\sum_{i=1}^{n_r}\frac{\mu^2 |\tilde{\delta d}|_i^2}{(\sigma_i+\mu)^2}-\eta^2\right)^2,
 % \end{equation}
% where $\tilde{\delta d}=V^{\top}{\delta d}$ and $W(\mu)$ is a diagonal matrix defined as
% \begin{equation} \label{W}
%     W(\mu)=
%     \begin{bmatrix}
%     \frac{\mu}{\sigma_1+\mu} &                          &                          & \\
%                              & \frac{\mu}{\sigma_2+\mu} &                          & \\
%                              &                          & \ddots                   & \\
%                              &                          &                          & \frac{\mu}{\sigma_{n_r}+\mu}
%     \end{bmatrix}.
% \end{equation}

% \begin{equation}
%     \|e(\mu)\|_2^2=\bigg\Vert\left(\frac{1}{\mu}{S} {S}^T+{I}\right)^{-1}\delta d({\varepsilon})\bigg\Vert_2^2=\tau n_r \sigma^2,
% \end{equation}
% .

% Assuming the eigenvalue decomposition of ${S} {W}^{-1}{S}^T=V\Sigma V^T$, we need to solve the following root finding problem for $\mu$:
% \begin{equation}
%     \bigg\Vert\left(\frac{\sigma}{\mu}+1\right)^{-1}\tilde{\delta d}({\varepsilon})\bigg\Vert_2^2=\tau n_r \sigma^2
% \end{equation}
% where $\tilde{\delta d}({\varepsilon})=V^T{\delta d}({\varepsilon})$.

\subsection{The method of generalized cross validation}
Generalized cross-validation (GCV) is a widely used method for selecting the regularization parameter without requiring prior knowledge of the noise level. Statistically, GCV is based on a data prediction strategy: an optimal regularization parameter should provide good predictions of missing or unseen data values. The optimum parameter is defined as
\begin{equation} \label{mu_gcv}
 \mu =  \arg\min_{\mu} \phi^{\text{gcv}}(\mu)=\frac{\|\bold{r}_{\mu}\|_2^2}{\left[\text{Tr}(\bI-\bS{\bS}^{\top} \left({\bS} {\bS}^{\top}+\mu{\bI}\right)^{-1})\right]^2}.
\end{equation}
where Tr denotes the trace of a matrix. Using eigenvalue decomposition of $\bS\bS^{\top}$, $\phi^{\text{gcv}}(\mu)$ can be written as
\begin{align} \label{mu_gcv_II}
\phi^{\text{gcv}}(\mu)&=\frac{\|\bold{r}_{\mu}\|_2^2}{\left[\text{Tr}(\bI-\bS{\bS}^{\top} \left({\bS} {\bS}^{\top}+\mu{\bI}\right)^{-1})\right]^2} \\
&=\frac{\|\bold{r}_{\mu}\|_2^2}{\left[\text{Tr}\left(\frac{1}{\mu}{\bS} {\bS}^{\top}+{\bI}\right)^{-1}\right]^2}\\
&=\frac{\|\bold{r}_{\mu}\|_2^2}{\left[\sum_{i=1}^{n_r}\frac{1}{\sigma_i/\mu+1} \right]^2}.
\end{align}
While standard GCV has good asymptotic properties (it works well with large datasets), it can be unreliable for small or medium-sized datasets. In such cases, GCV often selects a regularization parameter $\mu$ that is too small, leading to an under-regularized solution that fits the noise rather than the underlying signal. To address this deficiency, a more stable extension known as robust GCV (RGCV) has been proposed \citep{Lukas_2006_RGCV}. The RGCV criterion selects $\mu$ as the minimizer of 
\begin{equation}\label{phi_rgcv}
  \phi^{\text{rgcv}}(\mu) = \left[ \gamma+(1-\gamma)\chi(\mu) \right]  \phi^{\text{gcv}}(\mu),
\end{equation}
where
\begin{align} 
 \chi(\mu) &= \text{Tr}\!\left( \left( \bS\bS^{\top} \big(\bS \bS^{\top}+\mu \bI\big)^{-1} \right)^2 \right), \\
 &= \sum_{i=1}^{n_r} \left(\frac{\sigma_i}{\sigma_i+\mu}\right)^2, 
\end{align}
and $0<\gamma\leq 1$ is the robustness parameter. Substituting into the RGCV definition yields
\begin{equation}\label{mu_rgcv}
     \phi^{\text{rgcv}}(\mu) =  \left[\gamma+(1-\gamma)\sum_{i=1}^{n_r} \left(\frac{\sigma_i}{\sigma_i+\mu}\right)^2\right]\frac{\|\boldsymbol{r}_{\mu}\|_2^2}{\left[\sum_{i=1}^{n_r}\tfrac{1}{\sigma_i/\mu+1} \right]^2}. 
\end{equation}
The behavior of $\phi^{\text{rgcv}}(\mu)$ highlights why RGCV is more stable than standard GCV. As $\mu \to 0$, each ratio $\tfrac{\sigma_i}{\sigma_i+\mu}$ approaches $1$, so $\chi(\mu)\approx n_r$ and the factor $\gamma+(1-\gamma)\chi(\mu)$ amplifies $\phi^{\text{rgcv}}(\mu)$. This prevents RGCV from selecting an overly small $\mu$, which is a typical failure mode of standard GCV. On the other hand, when $\mu \to \infty$, $\chi(\mu)\to 0$ and the factor reduces to $\gamma$, acting as a simple rescaling that does not change the minimizer. Thus, RGCV modifies the shape of the GCV curve mainly near $\mu=0$, reducing the risk of under-regularization. The robustness parameter $\gamma$ controls the strength of this modification: smaller $\gamma$ values produce stronger corrections that favor smoother solutions, while larger $\gamma$ values leave the selection closer to GCV. In practice, moderate choices such as $\gamma=0.3$ often yield a good trade-off between stability and adaptability \citep{Lukas_2012_PRGCV}.

\subsection{The method of residual whiteness principle}
The autocorrelation of the data residual is defined as
\begin{equation}
    [C_{rr}(\mu)]_l = \sum_i [\br_{\mu}]_i^*[\br_{\mu}]_{i+l},
\end{equation}
where $l$ is the autocorrelation lag and $a^*$ denotes the complex conjugate of $a$. The autocorrelation value at zero lag is equal to the Euclidean norm or energy of the residual:
\begin{equation}
    [C_{rr}(\mu)]_0 = \sum_i [\br_{\mu}]_i^* [\br_{\mu}]_{i}=\|\br_{\mu}\|_2^2.
\end{equation}
We see that the DP method relies solely on the noise information at zero lag, without taking into account the autocorrelation at other lags. However, an important prior assumption about ideal white noise is that its autocorrelation is zero at all nonzero lags and equal to its total energy at zero lag \citep{Robinson_2000_GSA}. By normalizing the autocorrelation function with respect to its zero-lag value, the result becomes independent of the noise level. This normalized autocorrelation is defined as \citep{Sheriff_1995_ESX}
\begin{equation}
  C_{rr}^{\text{nrm}}(\mu)=  \frac{C_{rr}(\mu)}{[C_{rr}(\mu)]_0}.
\end{equation}
Ideally, $C_{rr}^{\text{nrm}}(\mu)$ equals one at zero lag and zero at all other lags (the Kronecker delta function). The deviation of $C_{rr}^{\text{nrm}}(\mu)$ from the delta function is a measure of whiteness \citep{Almeida_2013_PEB}.  Based on this principle, an optimal value of $\mu$ can be estimated by minimizing the energy of this difference \citep{Lanza_2013_VID}:
\begin{equation} \label{mu_rwp}
 \mu=\arg\min_{\mu} \phi^{\text{rwp}}(\mu)=\sum_{l=0}^{n-1}|[C_{rr}^{\text{nrm}}(\mu)]_l-\boldsymbol{\delta}_l|^2=  
\sum_{l=0}^{n-1} \Big\vert\frac{[C_{rr}(\mu)]_l}{\|\br_{\mu}\|_2^2}-\boldsymbol{\delta}_l\Big\vert^2,
\end{equation}
where $\boldsymbol{\delta}$ denotes the Kronecker delta function. 
This RWP not only eliminates the need for prior knowledge of the noise variance but also offers a potentially more accurate estimate of $\mu$, as it leverages richer statistical information from the residual.

According to the correlation theorem, the correlation between two signals is equal to the inverse Fourier transform of the product of the Fourier transform of one signal and the complex conjugate of the Fourier transform of the other \citep{Proakis_2001_DSP}. Let $F$ denote the discrete Fourier transform (DFT) matrix and $\hat{\br}_{\mu}=F\br_{\mu}$ be the Fourier transform of $\br_{\mu}$. Then, the autocorrelation function $C_{rr}(\mu)$ in the frequency domain is given by $|\hat{\br}_{\mu}|^2$:
\begin{equation}
 C_{rr}(\mu)   
 \underset{F^{-1}}{\overset{F}{\Longleftrightarrow}}
 \hat{\br}_{\mu}^* \circ \hat{\br}_{\mu}=|\hat{\br}_{\mu}|^2,
\end{equation}
where the magnitude and squaring are applied element-wise. Furthermore, by Parseval's theorem, the energy of a signal is preserved under the Fourier transform, i.e., $\|\br_{\mu}\|_2^2=\|\hat{\br}_{\mu}\|_2^2$. Based on these two theorems, the objective function in \eqref{mu_rwp} can be expressed more compactly in the Fourier domain as
\begin{align}
\phi^{\text{rwp}}(\mu)=\sum_{l=0}^{n-1} \Big\vert\frac{[C_{rr}(\mu)]_l}{\|\br_{\mu}\|_2^2}-\delta_l\Big\vert^2
&=\sum_{l=0}^{n-1}\Big\vert\frac{[F^{-1}|\hat{\br}_{\mu}|^2]_l}{\|\hat{\br}_{\mu}\|_2^2}-\delta_l\Big\vert^2\\
&=\sum_{l=0}^{n-1} \Big\vert F^{-1}\left(\frac{|\hat{\br}_{\mu}|^2}{\|\hat{r}_{\mu}\|_2^2}-F\delta\right)\Big\vert_l^2\\
 &=\sum_{l=0}^{n-1} \Big\vert \frac{|\hat{\br}_{\mu}|^2_l}{\|\hat{\br}_{\mu}\|_2^2}-1\Big\vert^2\\
  &=\sum_{l=0}^{n-1} \frac{|\hat{\br}_{\mu}|^4_l}{\|\hat{\br}_{\mu}\|_2^4}-2\sum_{l=0}^{n-1}\frac{|\hat{\br}_{\mu}|^2_l}{\|\hat{\br}_{\mu}\|_2^2}+1\\
  &=\frac{\|\hat{\br}_{\mu}\|_4^4}{\|\hat{\br}_{\mu}\|_2^4}-2+n.
\end{align}
Plugging this into \eqref{mu_rwp} and ignoring the constant terms, we get 
\begin{equation}\label{phi_rwp}
\phi^{\text{rwp}}(\mu)= \frac{\|\hat{\br}_{\mu}\|_4^4}{\|\hat{\br}_{\mu}\|_2^4}=\frac{\|F{\br}_{\mu}\|_4^4}{\|F{\br}_{\mu}\|_2^4}.
\end{equation}
We observe that the optimal value of $\mu$ is obtained by minimizing the ratio of the $\ell_4$-norm to the $\ell_2$-norm raised to the fourth power, a quantity known as kurtosis.

The concept of leveraging statistical properties, such as whiteness and kurtosis, has a long-standing and crucial role in geophysical signal analysis, particularly in deconvolution \citep{Robinson_2000_GSA}. This approach originated prominently with Minimum Entropy Deconvolution (MED), introduced by Wiggins \citep{Wiggins_1978_MED}.
The MED algorithm estimates an inverse filter that maximizes the kurtosis of the output signal, thereby producing a maximally spike-like or sparse reflectivity series. A similar concept underlies the proposed approach for regularization parameter selection. The RWP is based on the assumption that the data noise is additive, white, and Gaussian, implying that its autocorrelation function should resemble a spike—that is, a maximally sparse signal. Consequently, the RWP seeks to automatically select the penalty parameter ($\mu$) that maximizes the kurtosis of the residual autocorrelation. This maximization of autocorrelation ``spikiness” is equivalent to maximizing the whiteness of the residual, which can also be interpreted as minimizing the kurtosis of its Fourier-transformed counterpart.

\begin{figure}
 \center
 \includegraphics[width=0.75\columnwidth]{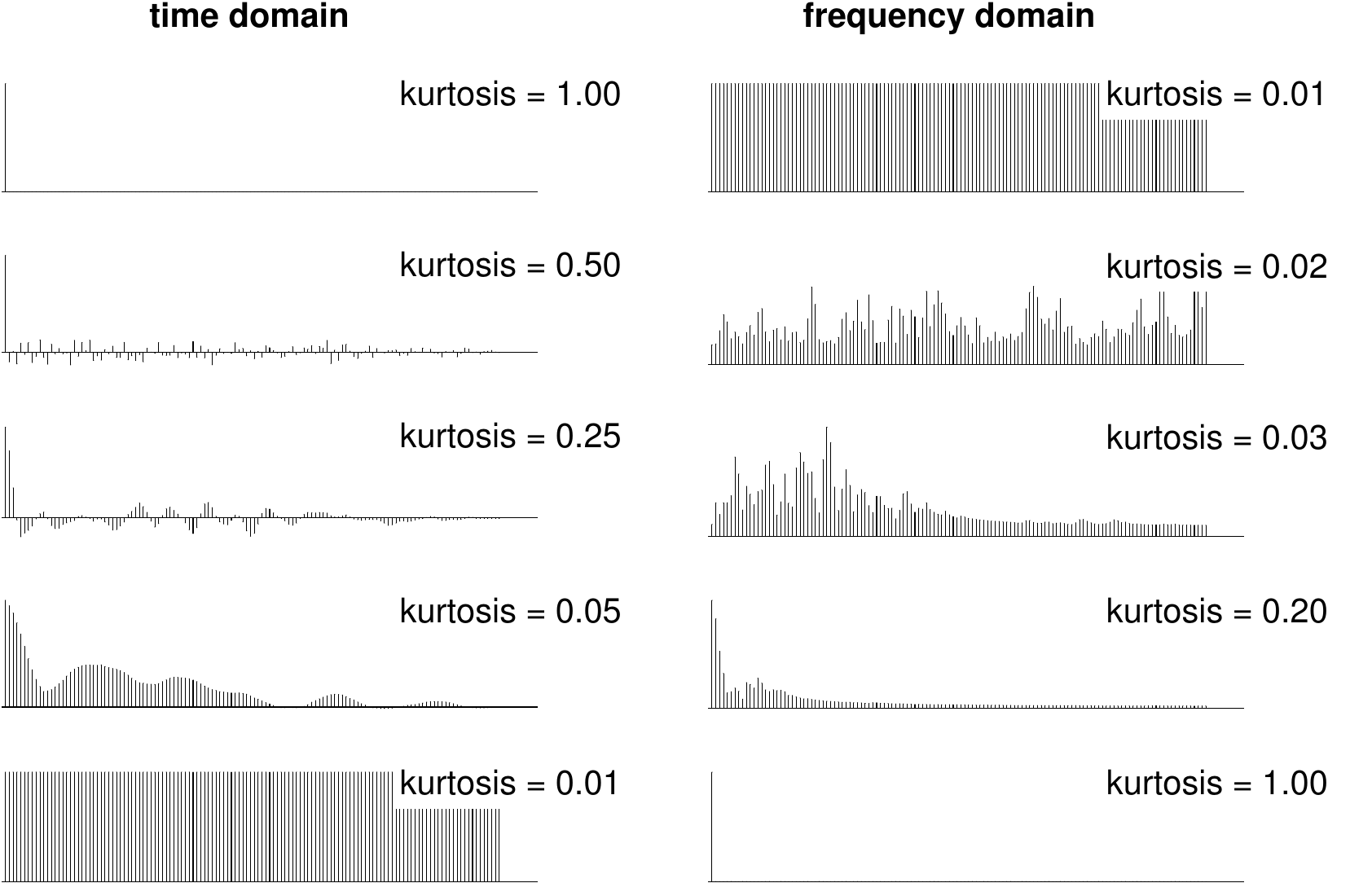}
 \caption{Examples of signals in the time domain (left) and their magnitude spectra in the frequency domain (right). The figure demonstrates the inverse relationship: signals transitioning from a sharp spike to a flat shape in the time domain (decreasing kurtosis) correspond to spectra evolving from flat to a spike in the frequency domain (increasing kurtosis). } 
 \label{fig:signals}
 \end{figure}

\fref{fig:signals} shows five different signals in both the time domain and frequency domain. From top to bottom, the time-domain signals transition from a spike to a flat shape. This corresponds to a monotonic decrease in their kurtosis values. In contrast, their corresponding frequency-domain spectra exhibit the opposite trend: the top signals show flat spectra (low kurtosis), while the bottom ones become increasingly concentrated at zero frequency, leading to higher kurtosis. This inverse relationship highlights how kurtosis can be used as a quantitative measure of flatness or spikeness in either domain, depending on the application context. In the context of penalty parameter estimation, we compute the predicted noise for various values of the penalty parameter and identify the optimal value as the one that yields the whitest residual. This residual whiteness can be assessed in two equivalent ways: by selecting the parameter that produces the most spiky autocorrelation function in the time domain (i.e., maximizing the Kurtosis), or the flattest power spectrum in the frequency domain (i.e., minimizing the Kurtosis).

%% %%%%%%%%%%%%%%%%%%%%%%%%%%%%%%%%%%%%%%%%%%%%%%%%%%%%%%%%%%%%%%%%%%%%%%%%%%%%%%%%%%%%%%%%%%%%%%%%
% Using the eigenvalue decomposition in \eref{eig}:
% \begin{equation} \label{phi_rwp}
% \phi^{\text{rwp}}(\mu)= \frac{\Vert F{V}W(\mu)\tilde{\delta d}\Vert_4^4}{\Vert F W(\mu)\tilde{\delta d}\Vert_2^4},
% \end{equation}
% where $V$ is the eigenvectors matrix in \eqref{eig} and $W(\mu)$ is defined in \eqref{W}.
% Once ${V}$ and $\tilde{\delta d}$ are computed, the global minimizer of \eqref{phi_rwp} can be efficiently determined at negligible computational cost.
%%%%%%%%%%%%%%%%%%%%%%%%%%%%%%%%%%%%%%%%%%%%%%%%%%%%%%%%%
%%%%%%%%%%%%%%%%%%%%%%%%%%%%%%%%%%%%%%%%%%%%%%%%%%%%%%%%%%
\subsection{Computational complexity of the parameter selection}
To locate the minimizer of each curve $\phi(\mu)$, we first evaluate $\phi(\mu)$ at $N$ logarithmically spaced $\mu$ values, identify the minimum (and if needed refine the search within a local interval around it). The main computational cost lies in the one-time eigenvalue decomposition of the $n_r\times n_r$ matrix $\bS\bS^\top$ ($\mathcal{O}(n_r^3)$), followed by a single projection step $\bold{V}^\top\delta\bd$ ($\mathcal{O}(n_r^2)$).
Once these are precomputed, evaluating $\phi(\mu)$ over $N$ candidates is inexpensive: $\mathcal{O}(N n_r)$ for DP and GCV, and $\mathcal{O}(N n_r + N n_r\log n_r)$ for RWP due to an additional Fourier transform. Therefore, since usually $N\leq n_r$, the total evaluation cost for all three methods scales as $\mathcal{O}(n_r^3)$, making parameter selection efficient even over large $\mu$ grids. Compared with the cost of a single FWI iteration—which requires solving a PDE—this cost is negligible, allowing adaptive parameter selection at each iteration with minimal overhead.

\begin{algorithm*}
 \begin{algorithmic}[1]
 \caption{Multiplier-based FWI (Augmented Lagrangian): Frequency domain implementation.} \label{alg1}
 \REQUIRE Observed data $\bd$, source $\bb$, initial model $\bold{m}$
 \FOR{$\omega\in[\omega_{min},\omega_{max}]$}
 \STATE  Set $\Eps^0_s=0$
  %------------------------------------------------------------------------
  %----------------------- inner loop -------------------------------------
  \FOR{$k=1,2,\texttt{maxit}$}
 \STATE Compute LU factorization of the wave-equation operator $\bold{A}^k\equiv \bA(\m^k)$ at the frequency $\omega$
 %%%%%%%%%%%%%%%%%%%%%%%%%%%%%%%%%%%%%%%%%%%%%%%%%%%%%%%%%%%%%%%%%%
  \FOR{$s=1$ to $n_s$}
  %-----------------------------------------------------------
 \STATE  \begin{tabular}{@{}p{6cm} p{10cm}@{}}
 Form the data-space Hessian matrix: & $\bH_{s}^k=\bold{S}_{s}^k(\bold{S}_{s}^k)^{\top}=\bP_s(\bA^k)^{-1}(\bA^k)^{-\top}\bP_s^{\top}$\\
  \end{tabular}
    %-----------------------------------------------------------
 % \STATE  \begin{tabular}{@{}p{6cm} p{10cm}@{}}
 % Compute the eigenvalues and eigenvectors of $H_{sk}$: & $H_{sk}=V_{sk}\Sigma_{sk} V_{sk}^T$\\
 %  \end{tabular}
 %-----------------------------------------------------------
 \STATE  \begin{tabular}{@{}p{6cm} p{10cm}@{}}
 Compute the data residuals: & $\delta\bold{d}^{k}_s=\bold{d}_s-\bold{P}_s(\bA^k)^{-1}(\bold{b}_s-\Eps^{k}_s)$\\
  \end{tabular}
 %-----------------------------------------------------------
 % {\color{red}
 % \STATE  \begin{tabular}{@{}p{6cm} p{10cm}@{}}
 % Compute $\mu^{k}_s$ (see \eqref{phi_rwp}): & $ \mu_s^k=\argmin_{\mu}  \phi^{\text{rwp}}(\mu),$ \\
 % \end{tabular}
 % }
% \STATE  \begin{tabular}{@{}p{6cm} p{10cm}@{}}
% \multirow{3}{6cm}{Compute $\mu^{k}_s$:} & 
% DP:~~~~ $ \mu_s^k=\underset{\mu}{\argmin}  \phi^{\text{dp}}(\mu)$ (see \eqref{mu_dp}) \\
% & GCV:~ $ \mu_s^k=\underset{\mu}{\argmin}  \phi^{\text{rgcv}}(\mu)$ (see \eqref{mu_rgcv}) \\
% & RWP: $ \mu_s^k=\underset{\mu}{\argmin}  \phi^{\text{rwp}}(\mu)$ (see \eqref{mu_rwp}--\eqref{phi_rwp} \\
% \end{tabular}
\STATE  \begin{tabular}{@{}p{6cm} p{10cm}@{}}
Compute $\mu^{k}_s$: & $\left\{
\begin{array}{l}
\text{DP: }~~~ \mu_s^k=\underset{\mu}{\argmin}  \phi^{\text{dp}}(\mu) \text{ (see \eqref{mu_dp})}\\
\text{RGCV: } \mu_s^k=\underset{\mu}{\argmin}  \phi^{\text{rgcv}}(\mu) \text{ (see \eqref{mu_rgcv})}\\
\text{RWP: } \mu_s^k=\underset{\mu}{\argmin}  \phi^{\text{rwp}}(\mu) \text{ (see \eqref{mu_rwp}--\eqref{phi_rwp})}
\end{array}
\right.$ \\
\end{tabular}
 %-----------------------------------------------------------
  \STATE \begin{tabular}{@{}p{6cm} p{10cm}@{}}
  Compute the backpropagating wavefield: &  
$ \sdual^{k+1}_s=(\bA^k)^{-\top}\bP_s^{\top}(\bold{H}_{s}^{k} +  \mu^k_s\bold{I})^{-1} \delta\bold{d}^k_s$\\
 \end{tabular}
  %-----------------------------------------------------------
\STATE \begin{tabular}{@{}p{6cm} p{10cm}@{}}
Compute the extended forward wavefield: &
$ \bold{u}^{k+1}_s =(\bA^k)^{-1}(\bold{b}_s +\sdual^{k+1}_s-\Eps^{k}_s)$ \\
 \end{tabular}
   %-----------------------------------------------------------
 \ENDFOR
  %%%%%%%%%%%%%%%%%%%%%%%%%%%%%%%%%%%%%%%%%%%%%%%%%%%%%%%%%%%%%%%%%%
\STATE \begin{tabular}{@{}p{6cm} p{10cm}@{}}
Update the model &
${\m}^{k+1}={\m}^{k}-\dfrac{\sum_{s=1}^{n_s}\langle \bL({\bu^{k+1}_s}), \sdual_s^{k+1}\rangle }{\sum_{s=1}^{n_s} \bL({\bu^{k+1}_s})^\top \bL({\bu^{k+1}_s})}$\\
 \end{tabular}
\STATE \begin{tabular}{@{}p{6cm} p{10cm}@{}}
Update the Lagrange multiplier: &
$\Eps^{k+1}_s=\Eps^{k}_s + \bold{A}(\bold{m}^{k+1})\bold{u}_s^{k+1}-\bold{b}_s$\\
 \end{tabular}
\ENDFOR
\STATE Use the model obtained at the current frequency, ${\m}^\texttt{maxit}$, as the initial model for the inversion at the next higher frequency.
\ENDFOR
\end{algorithmic}
\end{algorithm*}

%-----------------------------------------------------------------------------------
%--------------------------------------------- algorithm
%-----------------------------------------------------------------------------------
%%%%%%%%%%%%%%%%%%%%%%%%%%%%%%%%%%%%%%%% ALGORITHM 2 %%%%%%%%%%%%%%%%%%%%%%%%%%%%%%%%%%%%%%%%%%%%%
\begin{algorithm*}
 \begin{algorithmic}[1]
 \caption{Multiplier-based FWI (Dual Augmented Lagrangian): Frequency domain implementation.} \label{alg2}
 \REQUIRE Observed data $\bold{d}$, source $\bold{b}$, initial model $\m$
 \FOR{$\omega\in[\omega_{min},\omega_{max}]$}
 \STATE  Set ${\Eps}^0_s=0$
  %------------------------------------------------------------------------
  %----------------------- inner loop -------------------------------------
  \STATE Compute LU factorization of the wave-equation operator $\bold{A}^0\equiv\bold{A(m^0)}$ at the frequency $\omega$
  \STATE Compute Green functions for all receiver locations and build the Hessian matrices $\bold{H}_{s}^0=\bold{P}_s(\bold{A}^0)^{-1}(\bold{A}^0)^{-\top}\bold{P}_s^{\top}$ for $s=1,...,n_s$
  \FOR{$k=1$ to $\texttt{maxit}$}
 %%%%%%%%%%%%%%%%%%%%%%%%%%%%%%%%%%%%%%%%%%%%%%%%%%%%%%%%%%%%%%%%%%
  \FOR{$s=1$ to $n_s$}
 %-----------------------------------------------------------
 \STATE  \begin{tabular}{@{}p{6cm} p{10cm}@{}}
 Compute the data residuals: & $\delta\bold{d}^{k}_s=\bold{d}_s-\bold{P}_s(\bold{A}^0)^{-1}(\bold{b}_s-{\Eps}^{k}_s)$\\
  \end{tabular}
 %-----------------------------------------------------------
\STATE  \begin{tabular}{@{}p{6cm} p{10cm}@{}}
Compute $\mu^{k}_s$: & $\left\{
\begin{array}{l}
\text{DP: }~~~ \mu_s^k=\underset{\mu}{\argmin}  \phi^{\text{dp}}(\mu) \text{ (see \eqref{mu_dp})}\\
\text{RGCV: } \mu_s^k=\underset{\mu}{\argmin}  \phi^{\text{rgcv}}(\mu) \text{ (see \eqref{mu_rgcv})}\\
\text{RWP: } \mu_s^k=\underset{\mu}{\argmin}  \phi^{\text{rwp}}(\mu) \text{ (see \eqref{mu_rwp}--\eqref{phi_rwp})}
\end{array}
\right.$ \\
\end{tabular}
 %-----------------------------------------------------------
  \STATE \begin{tabular}{@{}p{6cm} p{10cm}@{}}
  Compute the backpropagating wavefield: &  
$ {\sdual}^{k+1}_s=(\bold{A}^0)^{-\top}\bold{P}_s^{\top}(\bold{H}_{s}^0 +  \mu^k_s\bold{I})^{-1} \delta\bold{d}^k_s$\\
 \end{tabular}
  %-----------------------------------------------------------
\STATE \begin{tabular}{@{}p{6cm} p{10cm}@{}}
Compute the extended forward wavefield: &
$ \bold{u}^{k+1}_s =(\bold{A}^0)^{-1}(\bold{b}_s +\sdual^{k+1}_s-{\Eps}^{k}_s)$ \\
 \end{tabular}
   %-----------------------------------------------------------
 \ENDFOR
  %%%%%%%%%%%%%%%%%%%%%%%%%%%%%%%%%%%%%%%%%%%%%%%%%%%%%%%%%%%%%%%%%%
\STATE \begin{tabular}{@{}p{6cm} p{10cm}@{}}
Compute the reflectivity model&
$\delta {\bold{m}}^{k+1}=- \dfrac{\sum_{s=1}^{n_s}\langle \bL({\bu^{k+1}_s}), \sdual_s^{k+1}\rangle }{\sum_{s=1}^{n_s} \bL({\bu^{k+1}_s})^\top \bL({\bu^{k+1}_s})}$\\
 \end{tabular}
\STATE \begin{tabular}{@{}p{6cm} p{10cm}@{}}
Update the Lagrange multiplier: &
${\Eps}^{k+1}_s={\Eps}^{k}_s + \bold{A}(\bold{m}^0+\delta \bold{m}^{k+1})\bold{u}_s^{k+1}-\bold{b}_s$\\
 \end{tabular}
\ENDFOR
\STATE Use the model obtained at the current frequency, ${\m}^\texttt{maxit}$, as the initial model for the inversion at the next higher frequency.
\ENDFOR
\end{algorithmic}
\end{algorithm*}

%% %%%%%%%%%%%%%%%%%%%%%%%%%%%%%%%%%%%%%%%%%%%%%%%%%%%%%%%%%%%%%%%%%%%%%%%%%%%%%%%%%%%%%%%%%%%%%%%%
 % \begin{figure}
 % \center
 % \includegraphics[width=1\columnwidth]{./noises}
 % \caption{??.}
 % \label{denoise}
 % \end{figure}

%% %%%%%%%%%%%%%%%%%%%%%%%%%%%%%%%%%%%%%%%%%%%%%%%%%%%%%%%%%%%%%%%%%%%%%%%%%%%%%%%%%%%%%%%%%%%%%%%%
 % \begin{figure}
 % \center
 % \includegraphics[width=0.5\columnwidth]{./denoise2}
 % \caption{(a) Noisy signal. (b) Denoised signal using first-order Tikhonov regularization with the RWP (blue) and DP (red), compared to the true signal (green). (c–d) Autocorrelation functions of the predicted noise using (c) RWP and (d) DP. (e) MSE, RWP \eqref{phi_rwp}, and DP \eqref{phi_dp} curves versus regularization parameter $\mu$; vertical dashed lines indicate the minimum location of each curve.} 
 % \label{denoise}
 % \end{figure}
%
%% %%%%%%%%%%%%%%%%%%%%%%%%%%%%%%%%%%%%%%%%%%%%%%%%%%%%%%%%%%%%%%%%%%%%%%%%%%%%%%%%%%%%%%%%%%%%%%%%
 \begin{figure}
 \center
 \includegraphics[width=0.8\columnwidth]{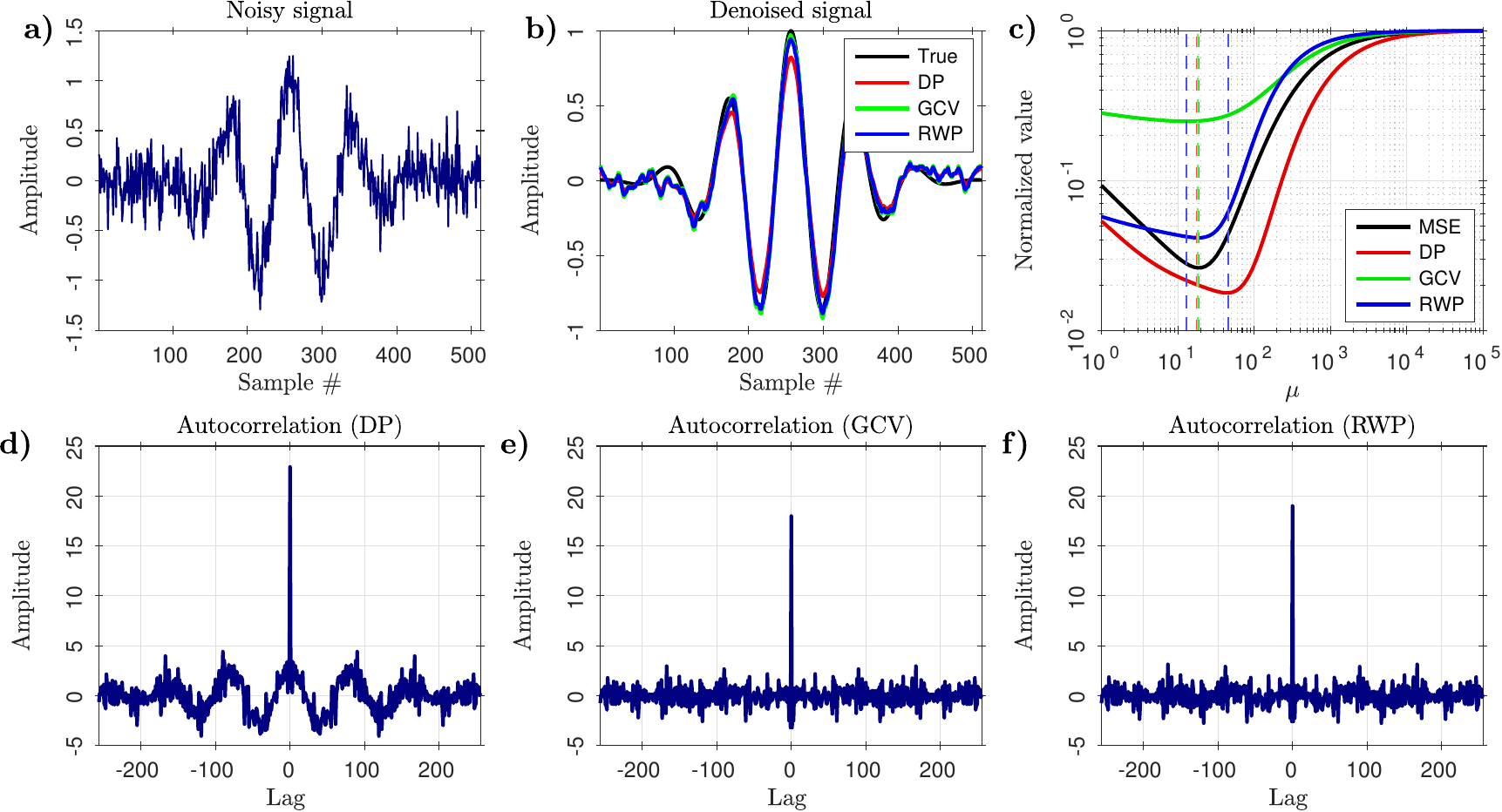}
 \caption{ Denoising Example: (a) Noisy Morlet wavelet. (b) Denoised wavelets using first-order Tikhonov regularization comparing DP, RGCV, and RWP methods. (c) RME, DP \eqref{mu_dp}, RGCV \eqref{phi_rgcv}, and RWP \eqref{phi_rwp} curves versus regularization parameter $\mu$; vertical dashed lines indicate the minimum location of each curve. (d–f) Autocorrelation functions of the predicted noise using (d) DP, (e) RGCV and (f) RWP.} 
 \label{denoise}
 \end{figure}
 
%  \vspace{-.65cm}

% \begin{table}[!h] 
% \caption{Computational performance of different FWI methods using DP, GCV, and RWP-based parameter selection strategies.}
% \label{tab:BP_runtime}
% \centering  
% \begin{tabular}{l c c c} 
% \toprule 
% & \multicolumn{3}{c}{Runtime (h)} \\ 
% \cmidrule(l){2-4} 
% Method & Reduced approach & Penalty approach & Dual-AL \\ 
% \midrule 
% DP   & 4.22 & 4.38 & 0.86 \\ 
% GCV  & 4.29 & 4.39 & 0.92 \\ 
% RWP  & 4.27 & 4.38 & 0.88 \\
% \bottomrule 
% \end{tabular}
% \end{table}

%
% \begin{figure*}
%     \centering
%     \includegraphics[width=1\textwidth]{figures/BPinv_DP_vs_RW_ver3.png}
%     \caption{ Comparison of inversion results under varying noise levels using DP (left column) and RWP methods (right column). Each row corresponds to a different noise level: (a) 20\%, (b) 30\%, and (c) 40\%. The adjacent velocity-depth profiles compare the estimated (red for DP and blue for RWP), initial (dashed gray), and true (solid gray) models. Reconstruction errors are indicated for each case.}
%     \label{fig:Inv_res}
% \end{figure*}

\section{Numerical examples}
In this section, we demonstrate the performance of the proposed parameter selection strategies—the DP, RGCV, and RWP—for both acoustic and elastic extended FWI on synthetic models. In our experiments, we benchmark our approach against the Reduced and Penalty formulations of FWI, which are considered special cases of the general AL framework (Algorithm \ref{alg1}). More specifically, setting the multiplier $\Eps_s$ to zero reduces the algorithm to the Penalty method \citep{VanLeeuwen_2013_MLM}. Furthermore, by setting $\Eps_s = 0$ and removing the source extension term $\sdual_s$ from the right-hand side of the wave equation, the algorithm simplifies to the Reduced approach, with the split Gauss-Newton Hessian \citep{Gholami_2024_FWI}.
For the GCV approach, we consistently employ the more stable Robust GCV (RGCV) criterion with $\gamma=0.3$ \citep{Lukas_2012_PRGCV}.

To assess reconstruction accuracy, we use the Relative Model Error (RME), defined as 
\begin{equation*}
    \text{RME} = \frac{\|\m^* - \m\|_2}{\|\m^*\|_2}.
\end{equation*}
Crucially, the DP is supplied with the exact noise level in this study, thereby removing the uncertainty associated with variance estimation and setting a high benchmark for comparison.

All computations were performed on a dual Intel Xeon Platinum 8176 processor system, comprising 56 cores operating at 2.10 GHz.
\subsection{A simple denoising example}
First, the performance of the different parameter selection methods is evaluated using a simple denoising example.  Figure~\ref{denoise}(a) shows a Morlet wavelet contaminated by additive Gaussian white noise. To denoise the signal, first-order Tikhonov regularization is employed, formulated as:
\begin{equation}
    \minimize_{\m}~ \frac12 \|\m-\bd\|_2^2+\frac{\mu}{2} \|\bD\m\|_2^2,
\end{equation}
where $\m$ denotes the desired denoised signal, $\bd$ is the noisy signal, $\bD$ is the finite-difference approximation of the first derivative operator, and $\mu>0$ is the regularization parameter. The DP, RGCV, and RWP are used to adaptively select $\mu$. Figure~\ref{denoise}(b) displays the denoised signals obtained using DP (red), RGCV (green), and RWP (blue) along with the true signal (black). The DP determines $\mu$ such that the norm of the residual matches the true noise level. However, due to the low-pass nature of Tikhonov filtering and signal-noise overlap at high frequencies, this often leads to signal attenuation. In contrast, RWP emphasizes the whiteness of the residual, leading it to select a smaller $\mu$ that better preserves signal features, even at the cost of retaining some noise. A similar behavior is observed for RGCV. This trade-off is evident in figure~\ref{denoise}(b), where the RWP and RGCV results more accurately match the true signal amplitudes between samples 100 and 400, albeit with increased noise elsewhere. As shown in figure~\ref{denoise}(c), this leads to a lower RME
for RWP compared to DP.
The autocorrelation functions of the residuals (predicted noise) are shown in figures~\ref{denoise}(d)--\ref{denoise}(f) for DP, RGCV, and RWP, respectively. The residuals from RWP and RGCV exhibit a sharp spike at zero lag and low-amplitude random values elsewhere, consistent with the characteristics of finite-length white noise. In contrast, the DP residual contains visible coherence at non-zero lags, indicating that some signal components have been misclassified as noise.

 \begin{figure}
    \centering
    \includegraphics[width=0.85\linewidth]{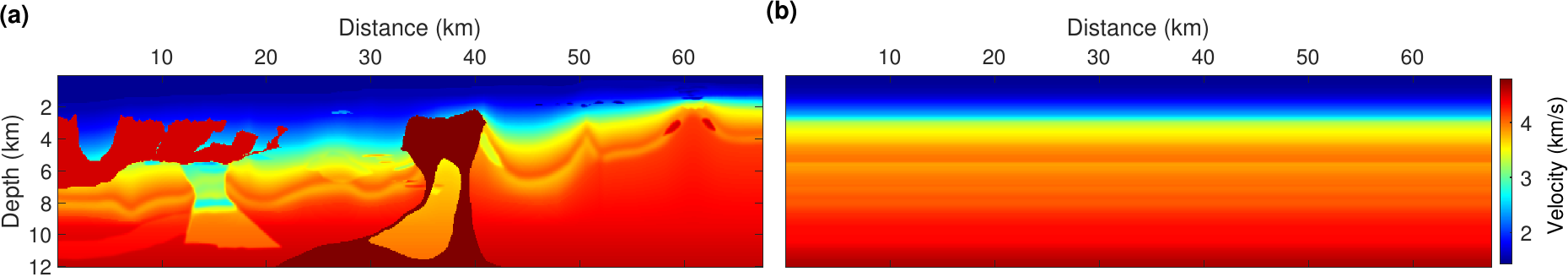}
    \caption{Acoustic FWI setup. (a) True 2004 BP velocity model. (b) Initial 1D velocity model used to start the inversion.}
    \label{fig:V_V0}
\end{figure}
\begin{figure}
    \centering
    \includegraphics[width=1\linewidth]{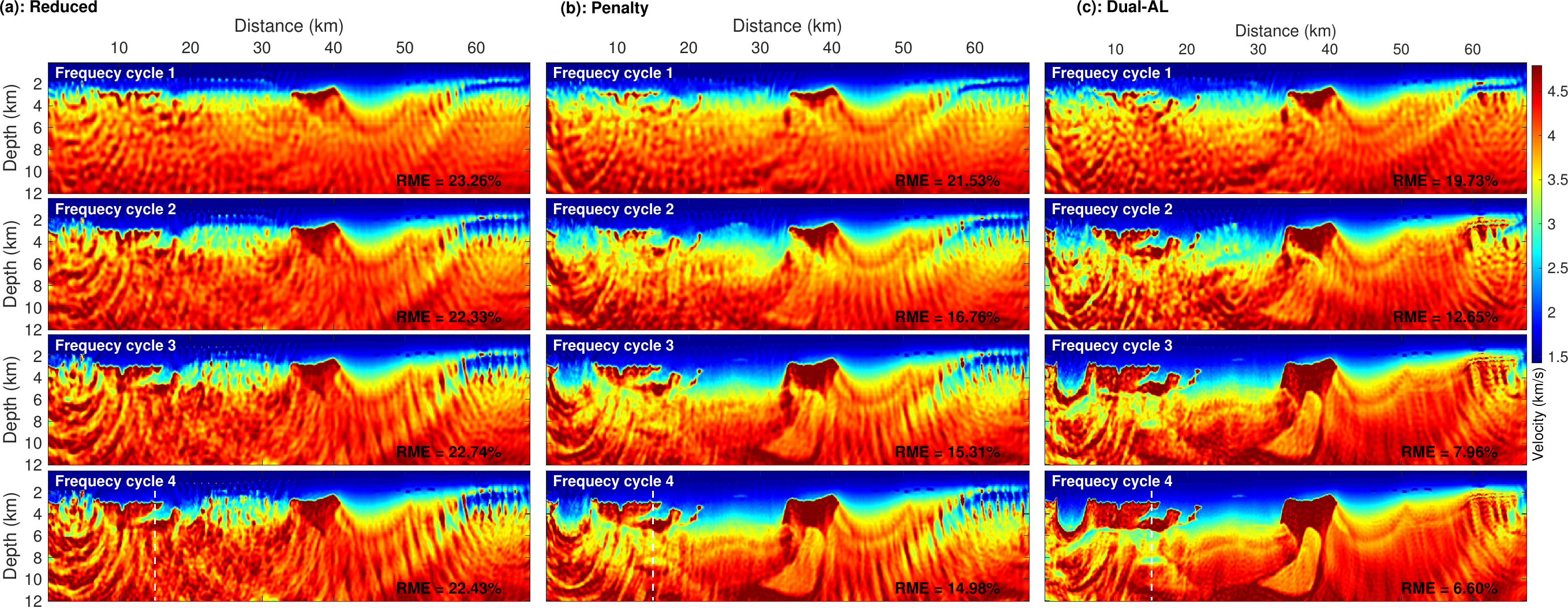}
    \caption{Noise-free acoustic inversion. Multiscale inversion results comparing the Reduced (a), Penalty (b), and Dual-AL (c) FWI methods after each frequency cycle, using the DP for penalty parameter selection.}
    \label{fig:BP_DP_noise_free}
\end{figure}
%
% \begin{figure}
%     \centering
%     \includegraphics[width=0.8\linewidth]{figures/BPinv_GCV_vs_RW_ver3.pdf}
%     \caption{2004 BP example (noise-free case). Final inversion results obtained by the reduced-space (a), penalty (b) and dual-AL (c) using GCV (left column) and RWP (right column)- based parameter selection strategies. Velocity profiles at $X = 14$ km (vertical white dashed lines) are compared with true and initial models, and the final RME (in percent) are reported. } 
%     \label{fig:BP_GCV_vs_RWP_noise_free}
% \end{figure}
%
  \begin{figure}  
  \centering     
  \includegraphics[width=1\linewidth]{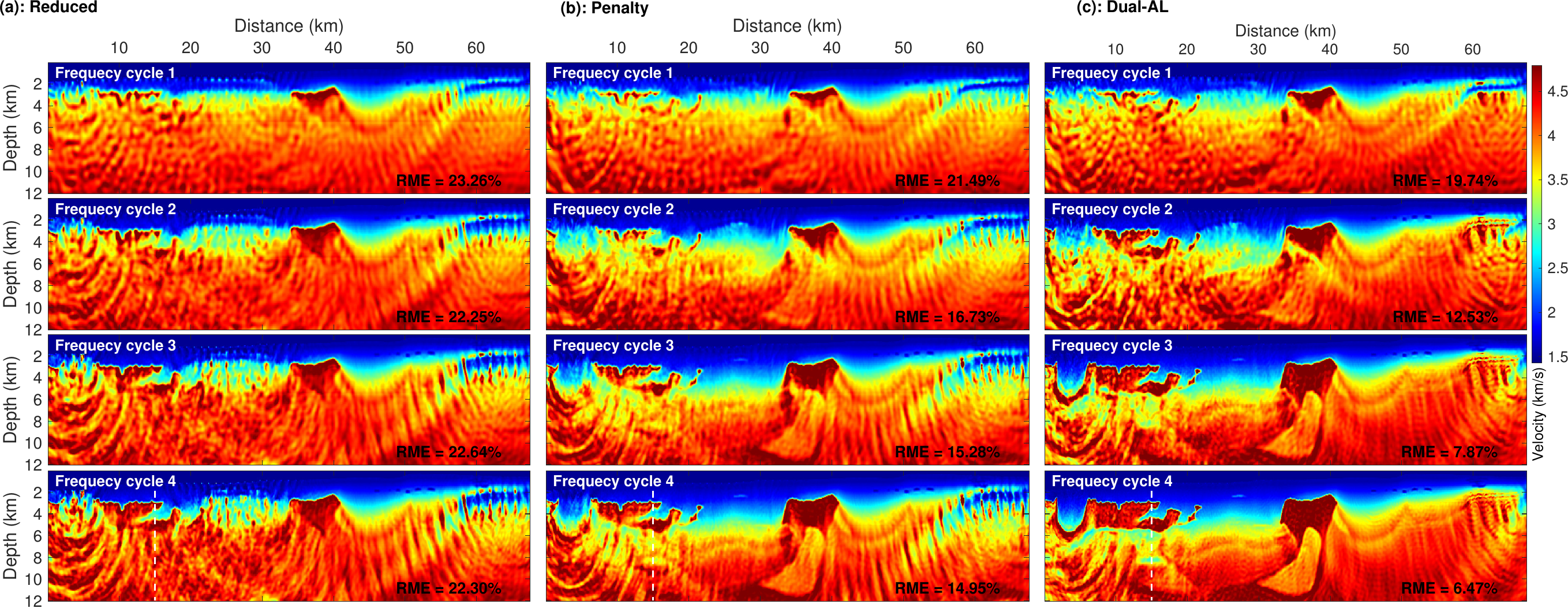}   
  \caption{Same as figure~\ref{fig:BP_DP_noise_free} but using the RGCV parameter selection.}  
  \label{fig:BP_GCV_noise_free}
  \end{figure}
  \begin{figure}  
  \centering     
  \includegraphics[width=1\linewidth]{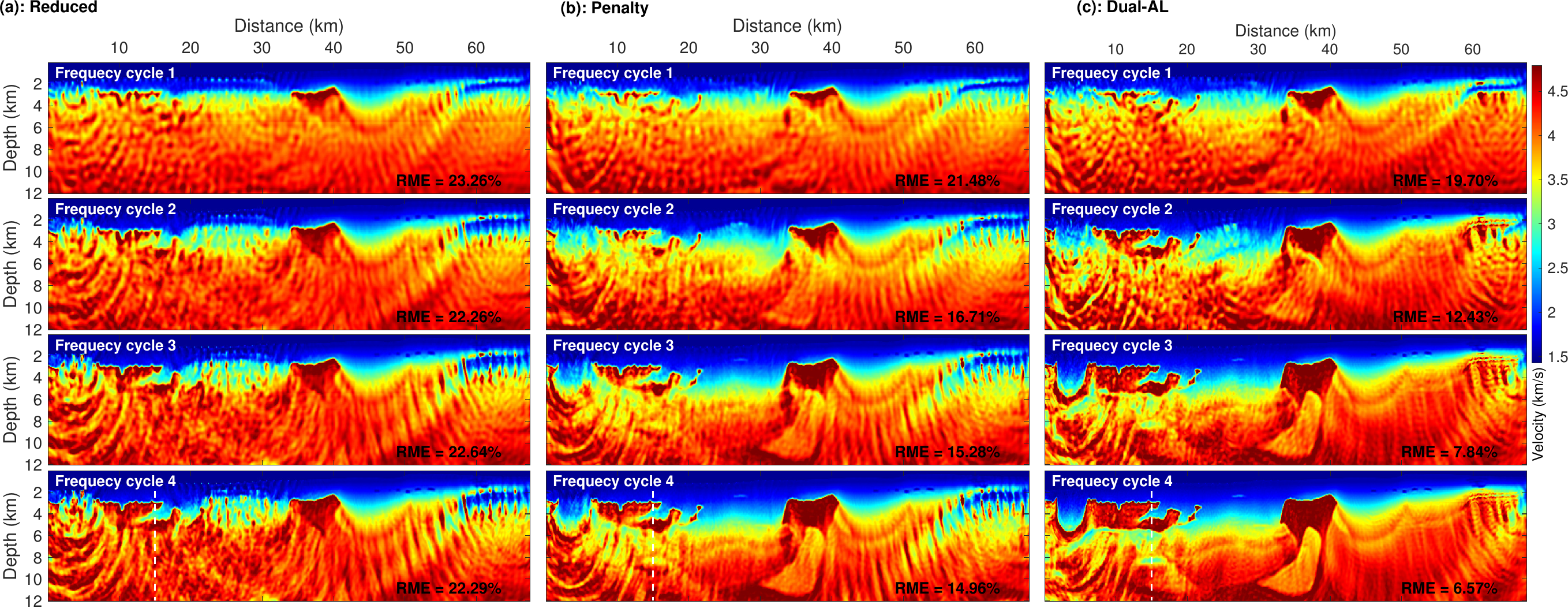}   
  \caption{Same as figure~\ref{fig:BP_DP_noise_free} but using the RWP parameter selection.}  
  \label{fig:BP_RWP_noise_free}
  \end{figure}

\begin{figure}  
  \centering     
  \includegraphics[scale=0.5]{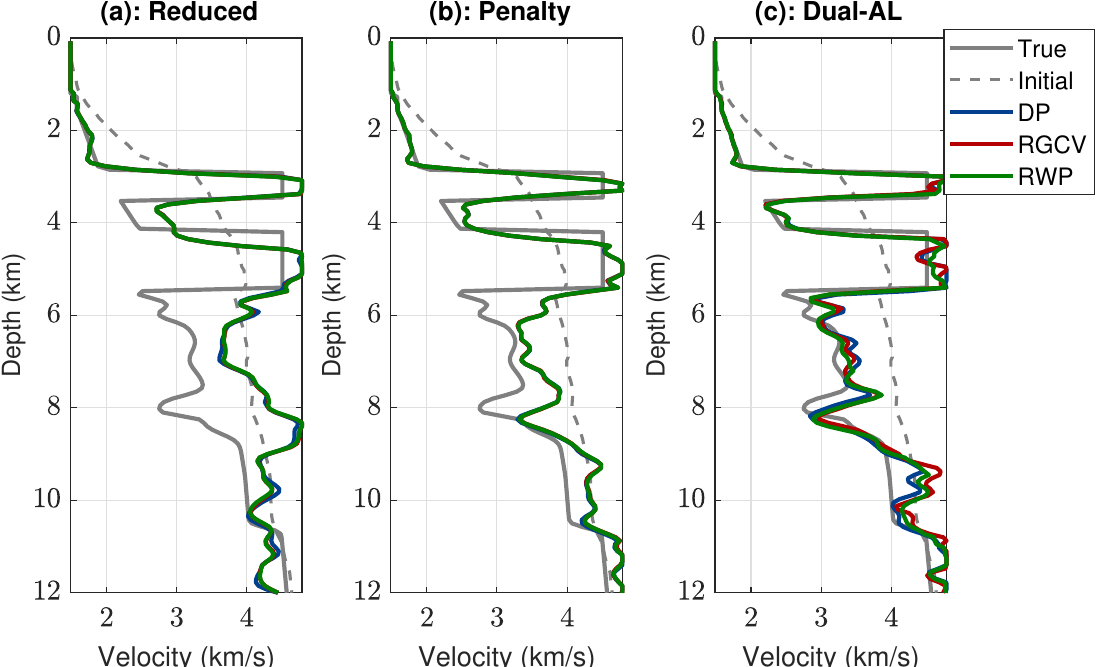}   
  \caption{Noise-Free acoustic inversion. Direct comparison between true, initial, and final estimated velocity profiles extracted at a 15 km distance. Results are shown for the (a) Reduced, (b) Penalty, and (c) Dual-AL FWI methods.} 
  \label{fig:BP_logs_noise_free}
\end{figure}

\subsection{Acoustic FWI example.}

Next, we evaluate the performance of different parameter selection strategies for the FWI algorithms described in Algorithms~\ref{alg1} and \ref{alg2}. To this end, we design a challenging synthetic experiment based on the 2004 BP velocity model, which spans $12~\mathrm{km} \times 67.5~\mathrm{km}$ with a grid spacing of 75~m, thereby preserving the original model dimensions (figure~\ref{fig:V_V0}(a)). The observed data are generated by simulating an ocean-bottom seismometer (OBS) acquisition system consisting of 67 seismometers deployed along the seafloor at 1~km intervals, together with 450 uniformly distributed pressure sources located at a depth of 75~m. A Ricker wavelet with a dominant frequency of 3~Hz is used as the source signature. To reduce computational cost, we exploit the spatial reciprocity of Green’s functions, enabling sources to be treated as receivers and vice versa.\\

For all experiments, inversion is initiated from the 1D starting model shown in figure~\ref{fig:V_V0}(b). This choice introduces significant challenges in the lack of low frequencies. In fact, \cite{Sun_2020_EFI} demonstrated that for similar initial model, cycle skipping occurs for inverting data frequencies above 0.3~Hz. To make the setup more realistic, we start the inversion from 1~Hz data and employ a multiscale inversion strategy \citep{Bunks_1995_MSW} over four frequency cycles:  
\begin{itemize}
    \item Cycle 1: 1--2.5~Hz, with 0.5~Hz intervals,  
    \item Cycle 2: 1--3~Hz, with 0.5~Hz intervals,  
    \item Cycle 3: 1--3.5~Hz, with 0.5~Hz intervals,  
    \item Cycle 4: 1--4.5~Hz, with 0.5~Hz intervals.  
\end{itemize}
In each case, the final model of one cycle serves as the initial model for the next. The water layer is assumed to be known during inversion.  

Two sets of experiments are carried out. In the first, we examine the performance of three FWI formulations— reduced, penalty, and dual-AL—under a noise-free scenario, using DP, RGCV, and RWP as parameter selection strategies. In the second, we focus on noisy data and evaluate the three parameter selection methods within the dual-AL framework against both random and colored noise.

\subsubsection{Computational efficiency and noise-free inversion.}
In the noise-free case, we first evaluate the computational efficiency and inversion performance of the Reduced, Penalty, and Dual-AL FWI formulations. The Reduced and Penalty formulations were set to a maximum of 385 iterations across the four frequency cycles, following traditional computational constraints. For the Dual-AL method, the number of inner iterations for updating the Lagrange multiplier ($\Eps$) was fixed at 10, which ensures convergence. For benchmarking, the DP was supplied with a tiny artificial noise level ($10^{-5}$).

The results summarized in Table~\ref{tab:BP_runtime} demonstrate the superior performance and dramatic computational efficiency gains provided by the Dual-AL method. The Reduced and Penalty methods required 385 LU factorizations and approximately 4.2 to 4.4 hours runtime. In stark contrast, the Dual-AL method required only 23 LU factorizations (a 94\% reduction), resulting in an approximate runtime of 0.9 hours (about 80\% faster). This efficiency is crucial because it validates the choice of Dual-AL as the necessary framework for testing computationally inexpensive, adaptive parameter selection strategies.
While all three FWI formulations showed consistent improvement across the four frequency cycles (figures~\ref{fig:BP_DP_noise_free}-\ref{fig:BP_RWP_noise_free}), the Dual-AL method delivered the best final performance, successfully recovering sharp geological boundaries, high-velocity zones, and subsalt features with minimal artifacts. This is reflected in the final RME values, where Dual-AL achieved the highest accuracy with RME values between 6.47\% and 6.60\%. This observation is further verified by the direct comparison of vertical velocity profiles at a 15 km distance (figure~\ref{fig:BP_logs_noise_free}), where the Dual-AL profiles provide the closest fit to the true model. Importantly, under noise-free conditions, the choice of parameter selection strategy (DP, RGCV, RWP) had a minimal impact on accuracy within any given method, with RME differences typically less than 0.1\%. RGCV showed slight superiority within the Dual-AL method (6.47\% RME), whereas DP performed the worst. This observation confirms that all regularization approaches converge to similar accurate solutions when data quality is high

\subsubsection{Inversion of noisy data: random noise.}
\begin{figure*}
    \centering
    \includegraphics[width=0.95\textwidth]{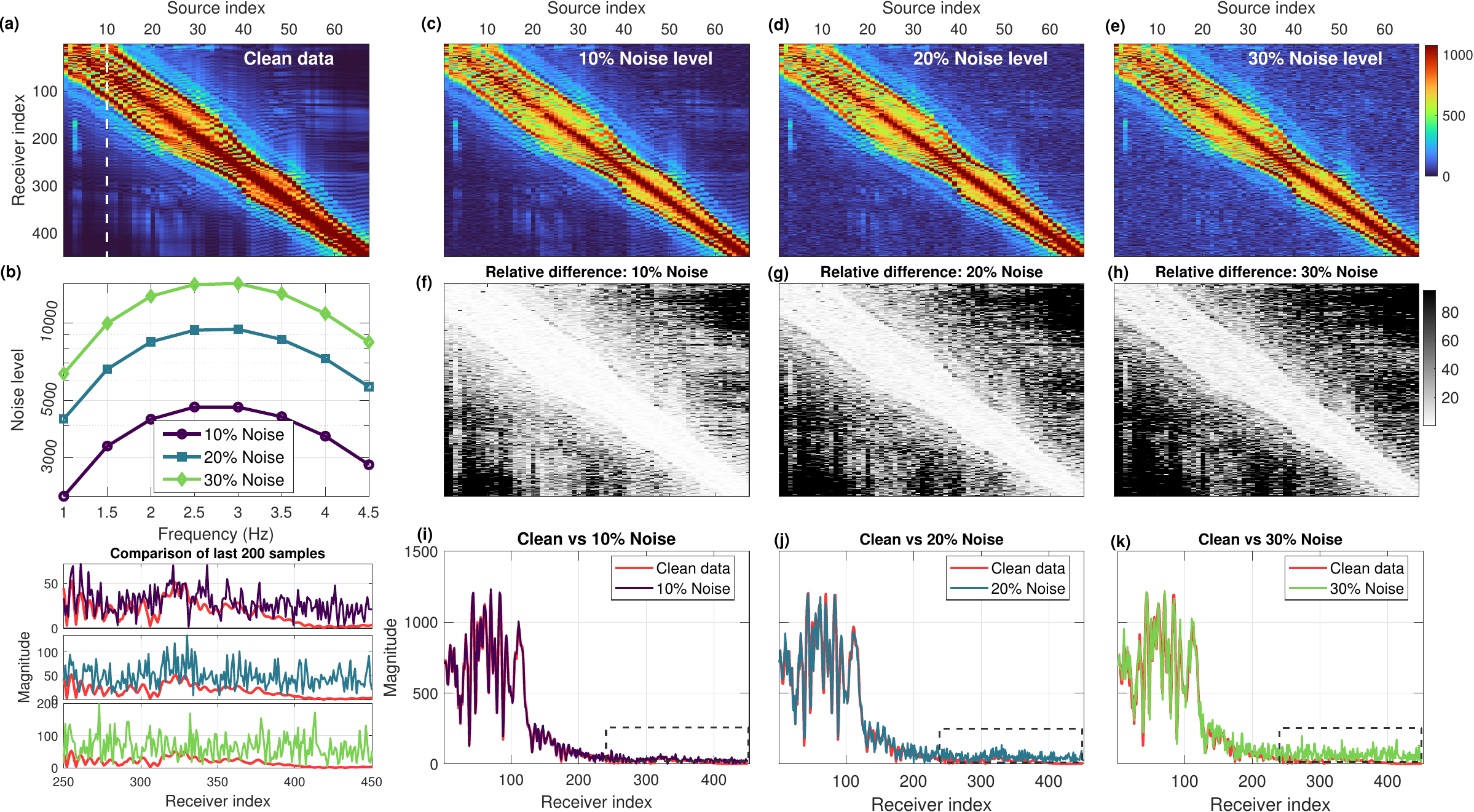}
    \caption{Impact of AWGN on data (frequency domain). Analysis of AWGN impact on monochromatic 2 Hz data (a). Panels (c–e) show noisy data at 10\%, 20\%, and 30\% noise levels, and the corresponding relative difference (f-h). Panels (i–k) compare magnitude, illustrating substantial amplitude variations at higher noise levels.
   % 2004 BP test. Analysis of the impact of additive Gaussian noise on data.  (a) Clean monochromatic 2 Hz data in shot-receiver coordinate. (b) Noise level versus frequency for the three cases.  (c-e) Noisy data with 10\%, 20\%, and 30\% noise levels, respectively. (f–h) Relative difference plots highlighting discrepancies between clean and noisy data for 10\%, 20\%, and 30\% noise levels. (i–k) Magnitude comparison between clean and noisy data at selected source (vertical dashed line in panel a) for 10\%, 20\%, and 30\% noise levels.  
    }
    \label{fig:data_Fourier}
\end{figure*}

\begin{figure*}
    \centering
    \includegraphics[scale=0.35]{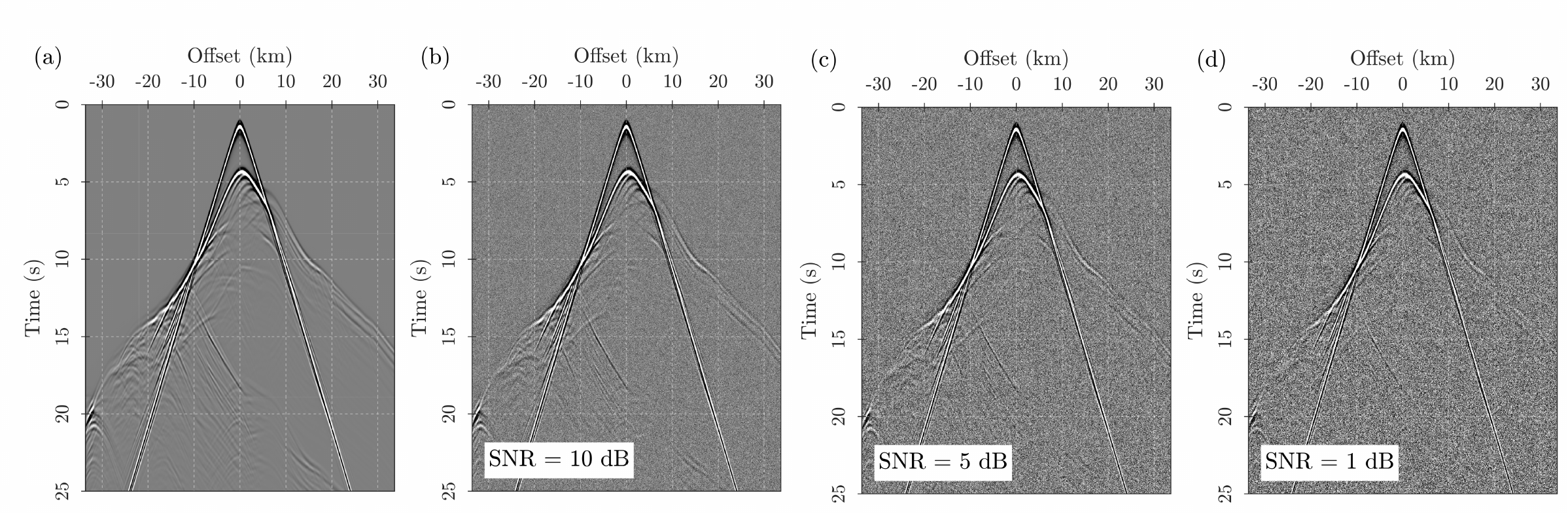}
    \caption{Impact of AWGN on data (time domain). (a) Noise-free reference data. (b–d) Data contaminated with 10\%, 20\%, and 30\% Gaussian noise, showing SNR reduction from 10 dB to 1 dB.
    }
    \label{fig:data_time}
\end{figure*}

We utilize the computationally efficient Dual-AL method as the baseline to investigate the robustness of DP, RGCV, and RWP under varying levels of Additive White Gaussian Noise (AWGN). Complex-valued AWGN is added to the data with levels set at 10\%, 20\%, and 30\% of the mean absolute value of the noise-free data.
Figures \ref{fig:data_Fourier} and \ref{fig:data_time} illustrate the severity of this noise. In the frequency domain (figure~\ref{fig:data_Fourier}), the monochromatic 2 Hz data shows a clear pattern that becomes progressively degraded with increasing noise, with substantial amplitude variations at 20\% and 30\% levels. In the time domain (figure~\ref{fig:data_time}), the signal-to-noise ratio decreases sharply from 10 dB at 10\% noise to a mere 1 dB at 30\% noise, making most seismic signals nearly indistinguishable from background noise

\begin{table}
\caption{Performance of different FWI methods using DP, RGCV, and RWP-based parameter selection strategies for 2004 BP model (noise-free case).}
\label{tab:BP_runtime}
\centering  
 \begin{adjustbox}{width=\textwidth}
\begin{tabular}{l c c c c c c c c c} 
\toprule 
& \multicolumn{3}{c}{Reduced} & \multicolumn{3}{c}{Penalty} & \multicolumn{3}{c}{Dual-AL} \\ 
\cmidrule(lr){2-4} \cmidrule(lr){5-7} \cmidrule(lr){8-10}
& LU.no & RME (\%) & Runtime (h) & LU.no & RME (\%) & Runtime (h) & LU.no & RME (\%) & Runtime (h) \\ 
\midrule 
Parameter selection & & & & & & & & & \\
\quad DP  & 385 & 22.43 & 4.22 & 385 & 14.98 & 4.38 & 23 & 6.60 & 0.86 \\
\quad RGCV & 385 & 22.30 & 4.27 & 385 & 14.95 & 4.38 & 23 & 6.47 & 0.88 \\
\quad RWP & 385 & 22.29 & 4.29 & 385 & 14.96 & 4.39 & 23 & 6.57 & 0.92 \\
\bottomrule 
\end{tabular}
 \end{adjustbox}
\end{table}

%
%
%%%%%%%%%%%%%%%%%%%%%%%%%%%%%%%%%%%%%%%%%%%%%%%%%%%%%%%%%%%%%%%%%%%
%
\begin{figure}
    \centering
    \includegraphics[width=0.85\linewidth]{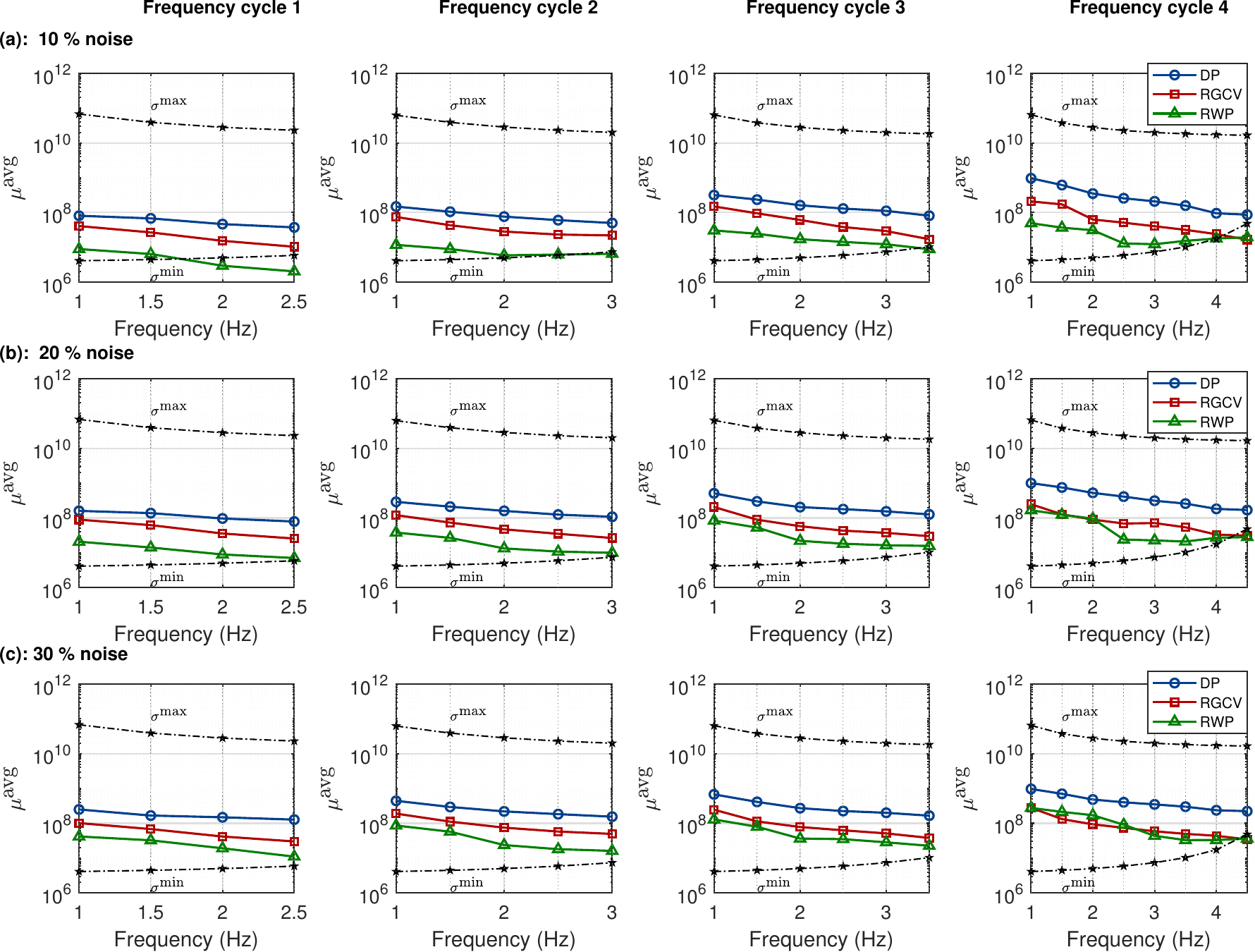}
    \caption{Average regularization parameter across frequency cycles for DP, RGCV, and RWP. The DP consistently selects $\mu$ values 1–2 orders of magnitude larger, while RWP and RGCV select smaller, more optimal parameters.
%    Average regularization parameter ($\mu^{\text{avg}}$) estimated using three parameter selection methods—DP, GCV, and RWP. The values are averaged over the inner iterations associated with the update of the multiplier for each test frequency and noise level. Results are shown for (a)~10\% noise, (b)~20\% noise, and (c)~30\% noise across four frequency cycles. The minimum and maximum singular values of $\bS\bS^{\top}$ ($\sigma^{\text{min}}$, $\sigma^{\text{max}}$) are also included.
    } 
    \label{fig:BP_noise_mu_vals}
\end{figure}
%%%%%%%%%%%%%%%%%%%%%%%%%%%%%%%%%%%%%%%%%%%%%%%%%%%%%%%%%%%%%%%%%%%%%
\begin{figure}
    \centering
    \includegraphics[width=.6\linewidth]{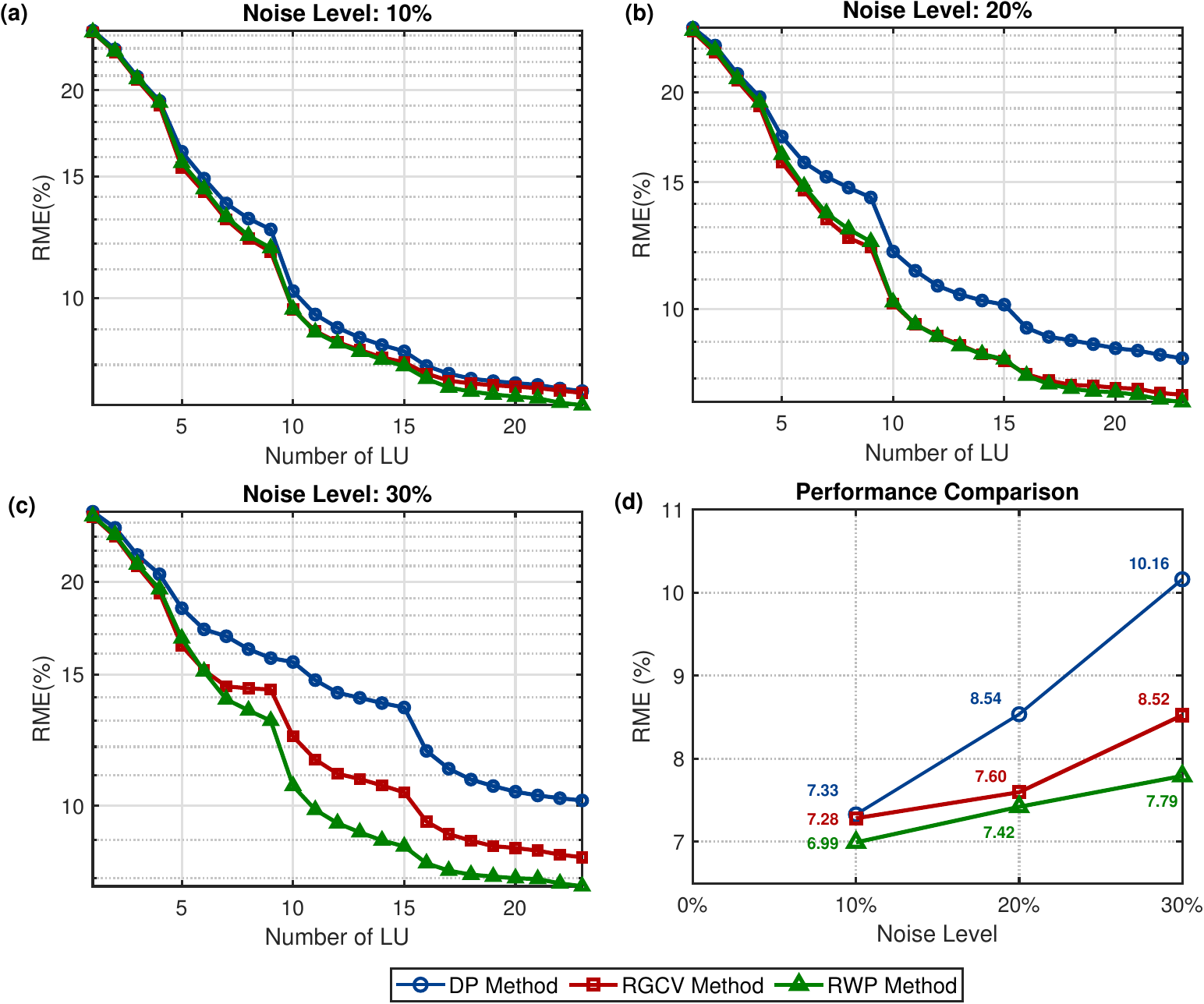}
    \caption{RME evolution versus number of LU factorizations for (a) 10\%, (b) 20\%, and (c) 30\% noise. Panel (d) summarizes the final RME versus noise level.
    % Quantitative comparison of DP, GCV and RWP methods under varying noise levels. Relative model error (in percent) versus the number of LU factorization for (a) 10\% noise, (b) 20\% noise, and (c) 30\% noise. (d) Performance comparison in terms of accuracy based on the final RME (in percent) versus noise level showing the behavior of each method.
    }
    \label{fig:BP_MSE_random}
\end{figure}
%%%%%%%%%%%%%%%%%%%%%%%%%%%%%%%%%%%%%%%%%%%%%%%%%%%%%%%%%%%%%%%%%%%%%%%
\begin{figure}
    \centering
    \includegraphics[width=1\linewidth]{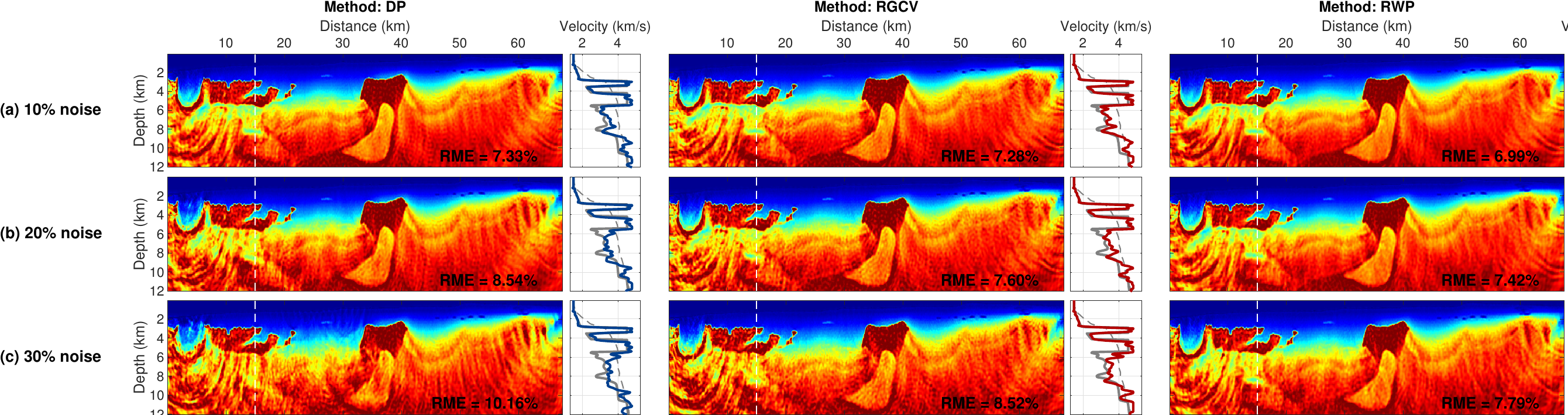}
    \caption{Final reconstructed models under AWGN. Final velocity models obtained using DP, RGCV, and RWP parameter selection methods (columns) under increasing noise levels: (a) 10\%, (b) 20\%, and (c) 30\% AWGN. Velocity profiles at a 15 km distance (vertical white dashed lines) are compared with true and initial models, and the final RME are reported.
    } 
    \label{fig:BP_inv_random_noise}
\end{figure}
%%%%%%%%%%%%%%%%%%%%%%%%%%%%%%%%%%%%%%%%%%%%%%%%%%%%%%%%%%%%%%%%%%%%%
\paragraph{Noise robustness and parameter selection philosophy.}
The central difference in performance stems from the parameter selection philosophy (figure~\ref{fig:BP_noise_mu_vals}), which directly impacts inversion accuracy (figure~\ref{fig:BP_MSE_random}).
From figure~\ref{fig:BP_MSE_random}(d), we see that the RWP exhibits exceptional noise robustness, maintaining a nearly constant RME between 6.99\% and 7.79\% across all noise levels tested (10\% to 30\%). This demonstrates that RWP consistently provides a superior regularization strategy for noisy data.
In contrast, the DP degrades significantly, increasing from 7.33\% RME at 10\% noise up to 10.16\% RME at 30\% noise, despite being supplied with the exact noise level. This degradation is reflected in the final reconstructed models, where DP exhibits notable artifacts and reduced resolution at high noise levels (figure~\ref{fig:BP_inv_random_noise}).

\begin{figure}
    \centering
    \includegraphics[scale=0.3]{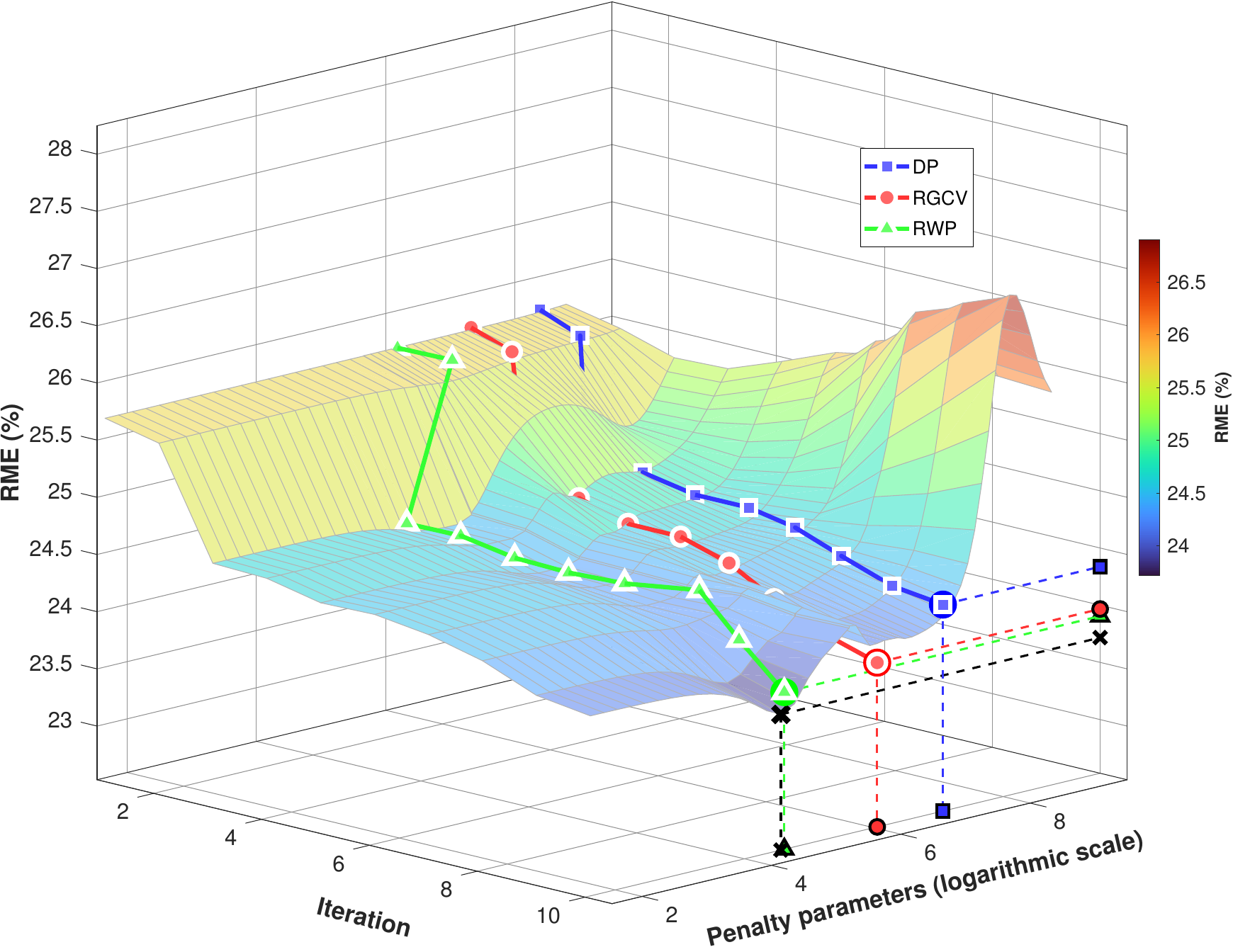}
    \caption{RME map on penalty parameter (10\% noise). RME values as a function of iteration number and fixed penalty parameter. Overlaid markers indicate the parameters automatically selected by DP, RGCV, and RWP, showing that RWP is closest to the global minimum RME (×).
   % RME values as a function of iteration number and penalty parameter $\mu$ for the inversion of 1~Hz data with 10\% noise. The surface plot shows the dependence of RME on $\mu$ across iterations, with overlaid markers indicating the automatically selected parameters from DP, GCV, and RWP methods.  The global minimum RME is marked by ×.
    }
    \label{fig:check_mu_fixed}
\end{figure}

\begin{figure}
    \centering
    \includegraphics[scale=0.4]{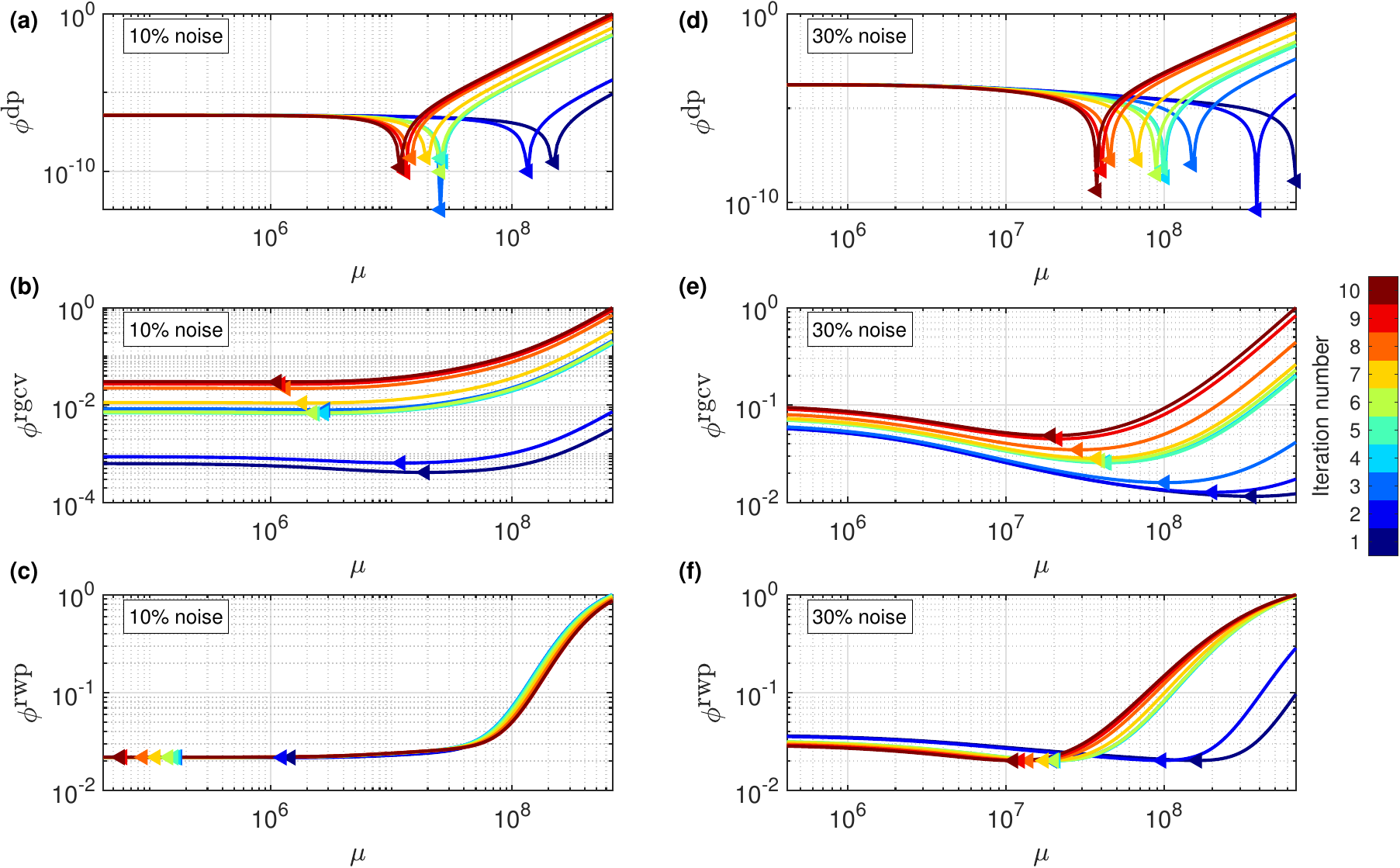}
    \caption{ Convergence of parameter objective functions. Evolution of $\phi^{\text{dp}}$, $\phi^{\text{rgcv}}$, and $\phi^{\text{rwp}}$ objective functions for 1 Hz data inversion across multiple iterations, comparing 10\% (a–c) and 30\% (d–f) noise levels. The plots show the adaptive increase in selected $\mu$ as noise increases.
    %Convergence behavior of DP, GCV, and RWP parameter selection methods under different noise conditions. (a-c) results for 10\% noise level. (d-f) results for 30\% noise level. Colors indicate iteration number according to the scale bar, with triangles marking the optimal parameter values selected by each method.
    } 
    \label{fig:phi_curves}
\end{figure}

This performance gap is explained by the parameter selection philosophy (figure~\ref{fig:BP_noise_mu_vals}). DP consistently selects $\mu$ values that are 1–2 orders of magnitude larger than RGCV and RWP, a gap that widens as noise increases. This aggressive regularization leads to overly smoothed, less accurate reconstructions because it sacrifices fine-scale features.
 RWP tends to favor smaller penalty parameters (figure~\ref{fig:BP_noise_mu_vals}). Analysis shows that the $\mu$ value selected by RWP almost reaches the global minimum RME (figure~\ref{fig:check_mu_fixed}), indicating that RWP provides the closest approximation to the optimal choice among the tested methods. This strategy, which emphasizes residual whiteness, better preserves weaker signals and low-wavenumber velocity components crucial for accurate reconstruction.
 The objective functions (figure~\ref{fig:phi_curves}) further confirm that all methods adaptively increase the selected $\mu$ as noise increases, and that the $\mu$ value decreases with iteration, highlighting the importance of the dynamic adjustment strategy introduced in this work.

%To further assess the ability of each method to identify optimal parameters, an additional experiment is performed for the case of inverting 1~Hz data with 10\% noise over 10 inner iterations of the dual-AL method. Here, the inversion is repeated using fixed $\mu$ values ranging from $10^{-5}\sigma_\text{min}$ to $\sigma_\text{max}$, logarithmically spaced into 50 samples, where $\sigma_\text{min}$ and $\sigma_\text{max}$ are minimum and maximum eigenvalues of $\bS \bS^{\top}$, respectively. The final RME values obtained during iterations for each fixed $\mu$ are plotted in Fig.~\ref{fig:check_mu_fixed}, together with the results of automatic parameter selection by DP, GCV, and RWP. The RWP method tends to favor smaller penalty parameters, achieving better data fitting, similar to the behavior noted in the denoising example in \fref{denoise}. Importantly, the $\mu$ value selected by RWP almost reach the global minimum RME (marked by ×), indicating that RWP provides the closest approximation to the optimal choice among the tested methods. The benefit of choosing smaller $\mu$ values lies in preserving weaker signals at long offsets, though at the cost of retaining some noise. Nevertheless, these preserved long-offset signals play a crucial role in recovering the low-wavenumber components of the velocity model. 
%
%%%%%%%%%%%%%%%%%%%%%%%%%%%%%%%%%%%%%%%%%%%%%%%%%%%%%%%%%%%%%%%%%%%%%%%
%%%%%%%%%%%%%%%%%%%%%%%%%%%%%%%%%%%%%%%%%%%%%%%%%%%%%%%%%%%%%%%%%%%%%%%%
%%%%%%%%%%%%%%%%%%%%%%%%%%%%%%%%%%%%%%%%%%%%%%%%%%%%%%%%%%%%%%%%%%%%%%%
\begin{figure}
    \centering
    \includegraphics[width=0.90\linewidth,trim={0cm 0cm 0cm 0cm},clip]{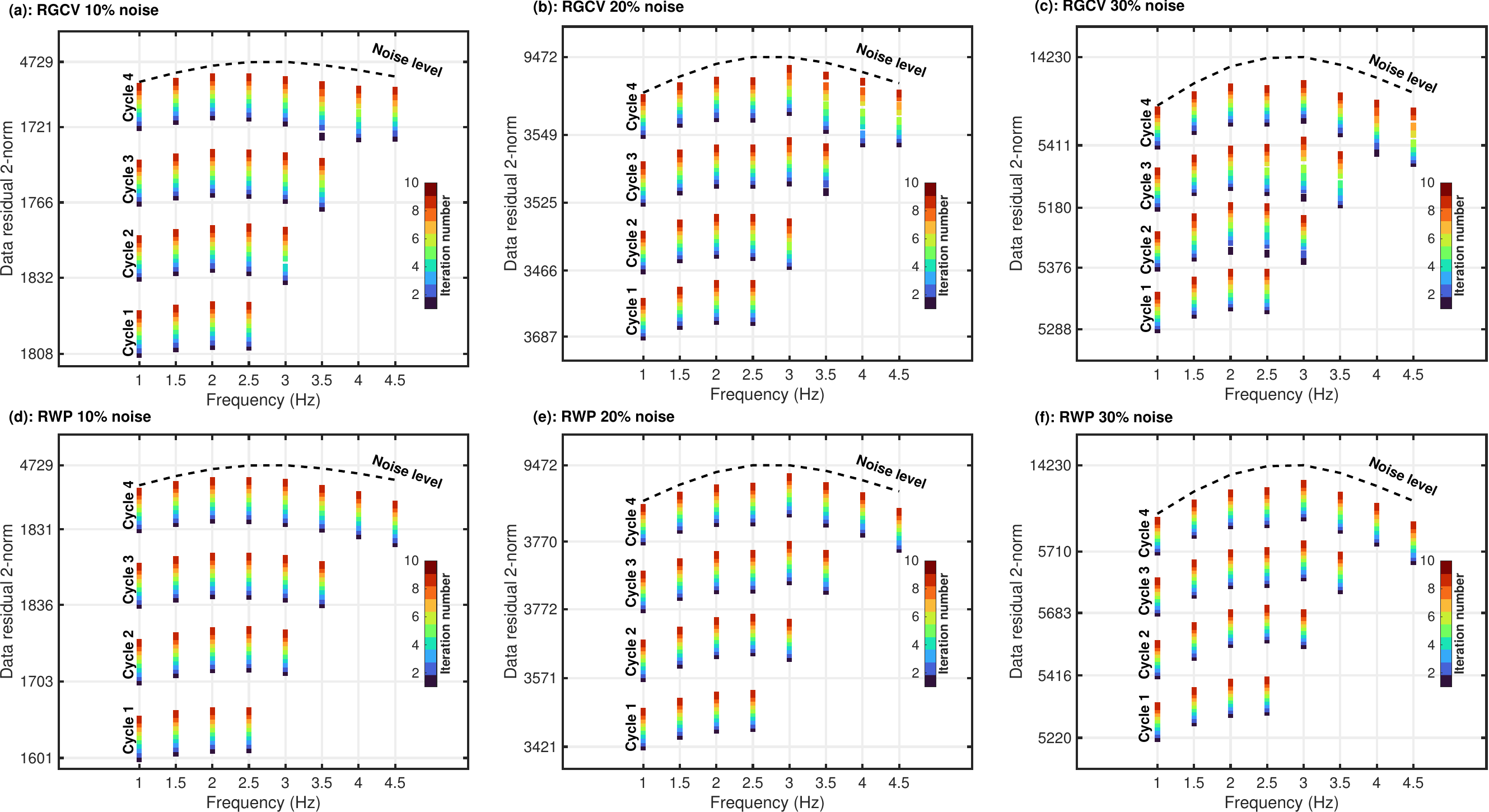}
    \caption{Evolution of data residual misfit (RGCV vs. RWP). Evolution of the 2-norm data residuals across frequencies and iterations for RGCV (top row) and RWP (bottom row) under 10\%, 20\%, and 30\% noise. The dashed curves indicate the true noise level, showing that both methods fit the data precisely at the noise level by the final cycle. 
  %  Evolution of 2-norm data residuals during iterations for different regularization parameter selection strategies across frequencies under varying noise conditions. The plots display data residual values versus frequency (1-4.5 Hz) across four frequency cycles comparing GCV (top row) and RWP (bottom row) methods under three noise levels: (a,d) 10\% noise, (b,e) 20\% noise, and (c,f) 30\% noise. Colors represent different iteration numbers during the inversion process, progressing from early iterations (blue) to later iterations (red) as indicated by the color bar (iteration number 1-10). The dashed curves indicate true noise levels at each frequency.
    }
    \label{fig:Data_Res_BP}
\end{figure}
%%%%%%%%%%%%%%%%%%%%%%%%%%%%%%%%%%%%%%%%%%%%%%%%%%%%%%%%%%%%%%%%%%%%%%%%%%
\paragraph{ Data fitting and convergence.}
The convergence behavior of RWP and RGCV demonstrates their adaptive approach to noise mitigation (figure~\ref{fig:Data_Res_BP}). Unlike DP, which forces the data misfit to match the noise level across all iterations, RWP and RGCV select smaller parameters during intermediate iterations, allowing for tighter data fitting and hence less wave-equation satisfaction. 
By the final iteration of the last frequency cycle, however, both RWP and RGCV fit the data precisely at the true noise level (indicated by the dashed line in figure~\ref{fig:Data_Res_BP}). The overall convergence curves (figure~\ref{fig:Data_Res_BP}) show that RWP consistently maintains a good fit even as noise increases.

\begin{figure*}
    \centering
    \includegraphics[width=1\linewidth]{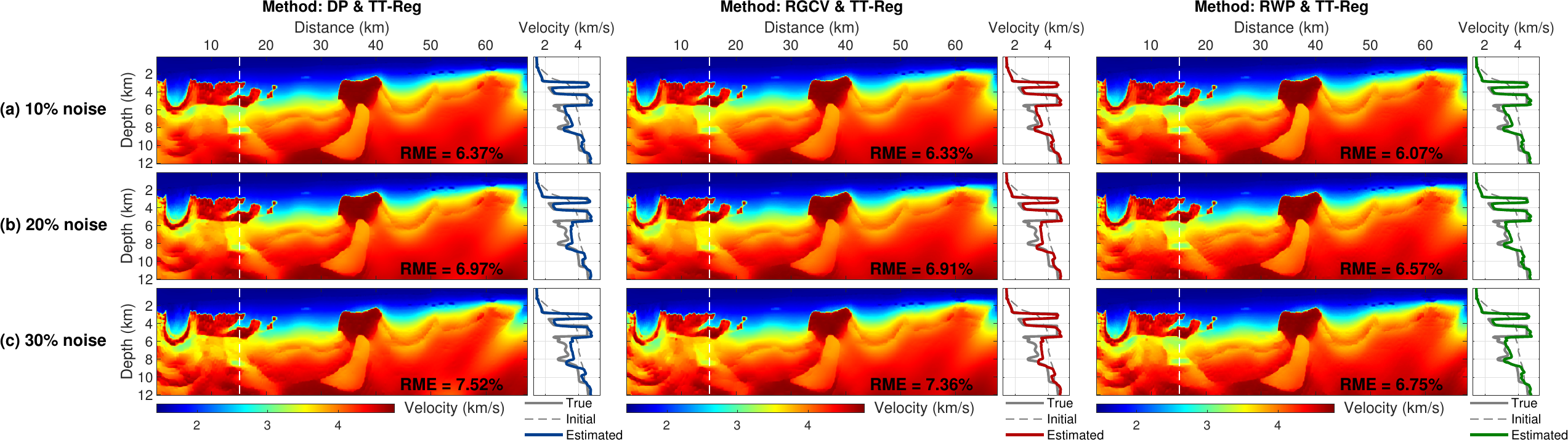}
    \caption{Final reconstructed velocity models using adaptive TT regularization, comparing DP, RGCV, and RWP under 10\%, 20\%, and 30\% noise.} 
    \label{fig:BP_REG}
\end{figure*}

\begin{figure*}
    \centering
    \includegraphics[width=0.7\linewidth]{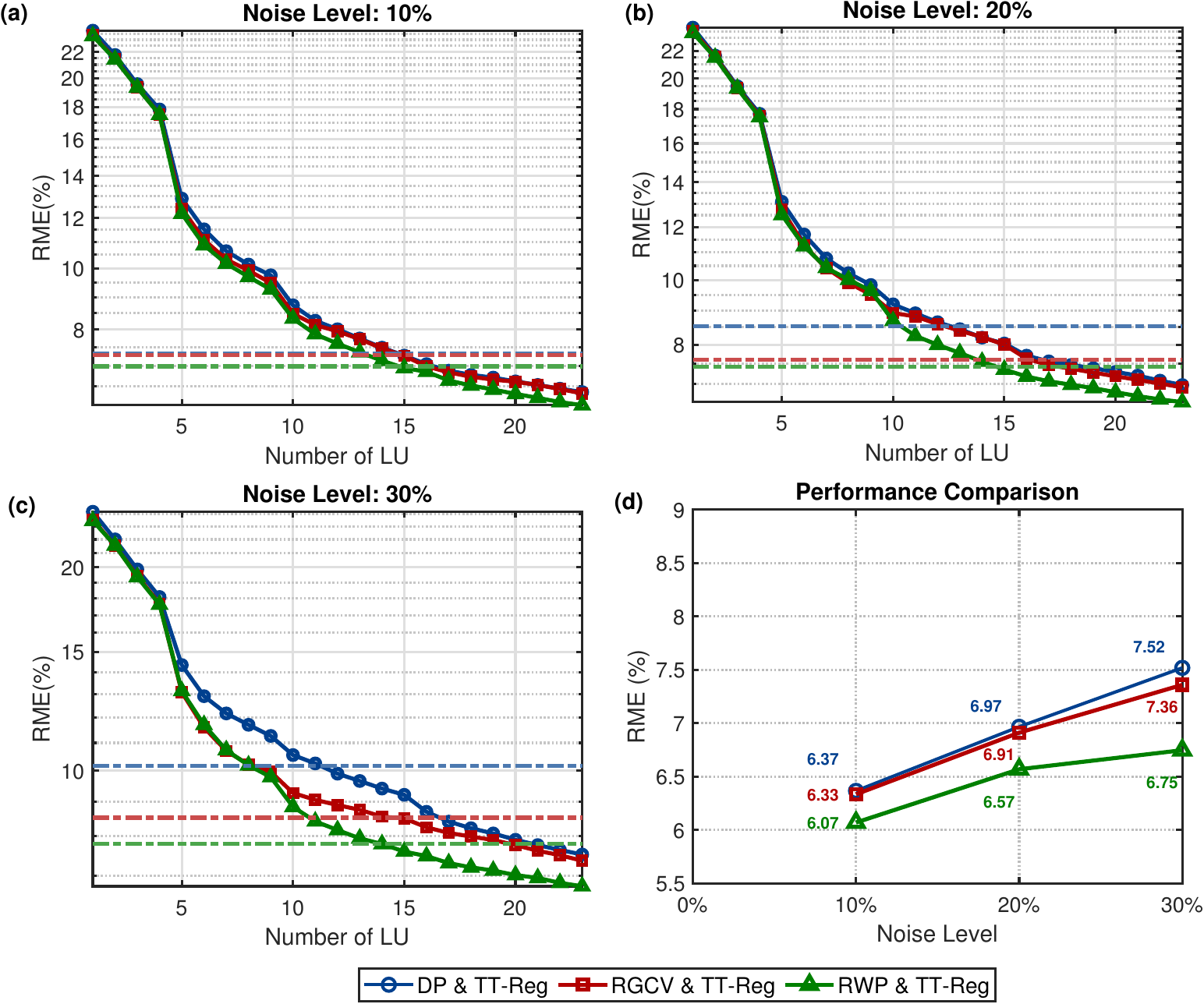}
    \caption{RME evolution versus LU factorizations for TT-regularized inversion under (a) 10\%, (b) 20\%, and (c) 30\% noise. Panel (d) shows that RWP + TT-Reg achieves the lowest RME across all noise levels. The horizontal dash-dot lines indicate the final RME level without regularization.
    } 
    \label{fig:BP_REG_MSE}
\end{figure*}

\paragraph{Comparison in the presence of regularization.}
The Dual-AL framework was tested with adaptive TT regularization  \citep{Gholami_2022_ABP,Aghazade_2025_RAE}.
TT regularization significantly improved reconstruction quality compared to the unregularized case (figure~\ref{fig:BP_inv_random_noise}), particularly in complex geological features and the sub-salt region (figure~\ref{fig:BP_REG}). When combined with this regularization, RWP consistently achieved the lowest RME values across all noise levels (figure~\ref{fig:BP_REG_MSE}), further demonstrating its robust selection capability.

%The proposed algorithm can be readily extended to include regularization in the objective function \eqref{main}, affecting only the model-update subproblems \eqref{ALM22} and \eqref{DALM22}. To regularize the model update, we employ adaptive Tikhonov–Total variation (TT) regularization \citep{Gholami_2022_ABP}. This approach decomposes the model into smooth and blocky components, allowing robust recovery of piecewise-smooth structures by applying Tikhonov and total-variation regularization separately to each part. Detailed implementation of TT regularization within the AL framework is beyond the scope of this work; interested readers are referred to \citep{Aghazade_2025_RAE}.

% \begin{figure*}
%     \centering
%     \includegraphics[width=0.9\linewidth]{figures/BP_RWP_REG.pdf}
%     \caption{2004 BP example (effect of regularization). Multiscale inversion results of dual-AL with RWP and adaptive TT regularization for 10\% (a), 20\% (b), and 30\% (c) noise levels across frequency cycles.} 
%     \label{fig:BP_REG}
% \end{figure*}

%

\begin{figure*}
    \centering
    \includegraphics[scale=0.55]{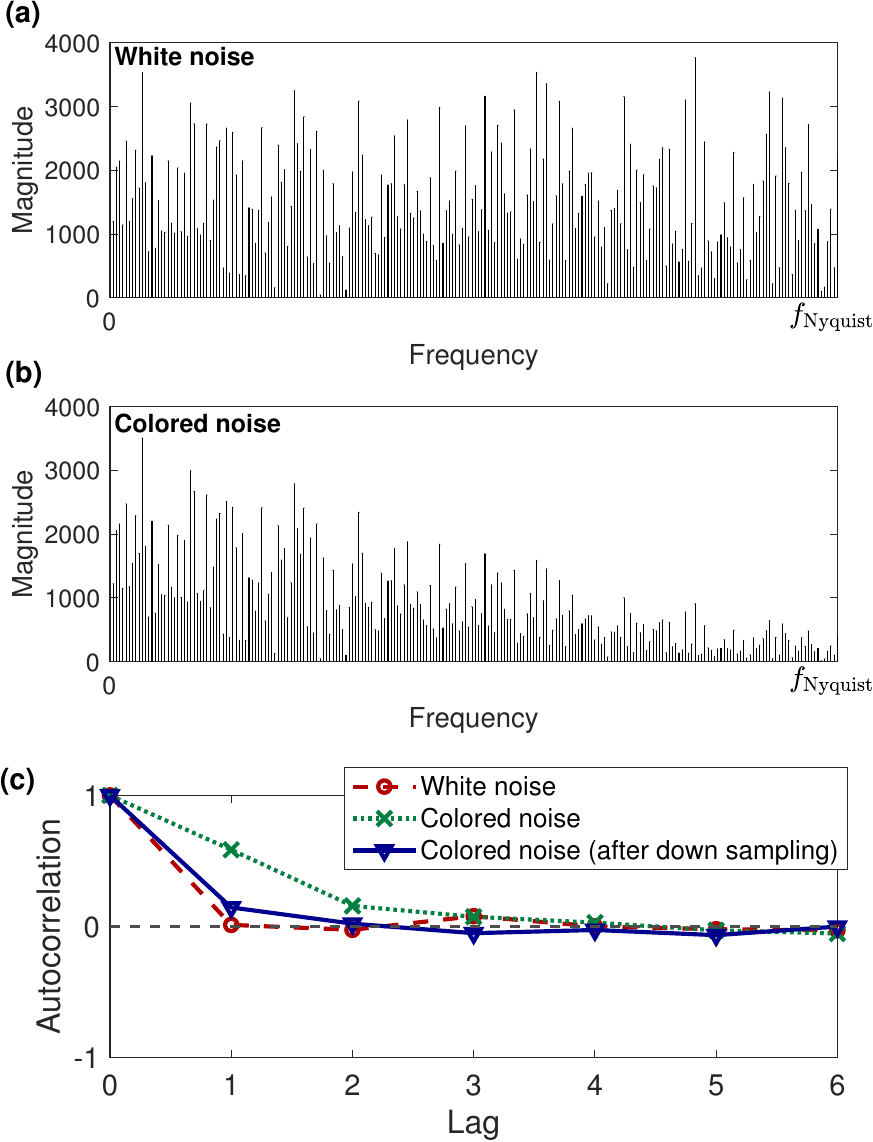}
    \caption{Colored noise analysis. Frequency-domain spectrum of (a) White noise and (b) Colored noise. (c) Autocorrelation functions showing that colored noise (green) exhibits non-negligible correlation at small lags, which is reduced via down-sampling (blue).}
    \label{fig:colored_noise}
\end{figure*}

\begin{figure*}
    \centering
    \includegraphics[scale=.5]{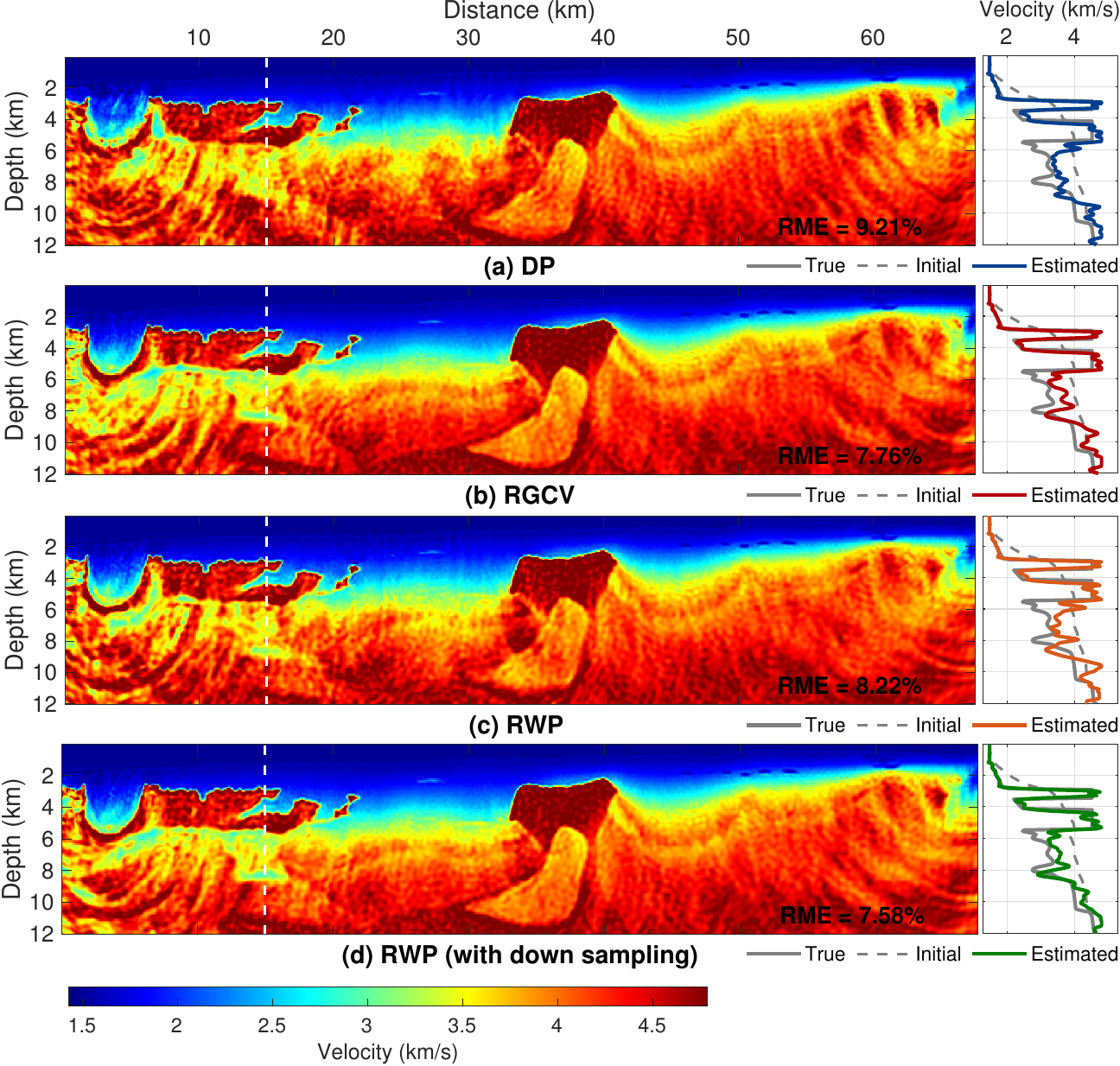}
    \caption{Final inversion results under colored noise. Comparison of final acoustic inversion results under colored noise for (a) DP, (b) RGCV, (c) RWP, and (d) RWP utilizing down-sampling. 
    Velocity-depth profiles on the right compare the true model (solid line), initial model (dashed line), and estimated model (colored line) at 15 km distance (shown by vertical white dashed line).}
    \label{fig:colored_BP}
\end{figure*}

% \begin{figure}
%     \centering
%     \includegraphics[width=.5\linewidth]{figures/MSE_BP_REG.pdf}
%     \caption{2004 BP example (effect of regularization). Quantitative comparison of the RWP method with and without TT regularization under varying noise levels. 
%     Panels (a)--(c) show the relative model error (RME, in percent) versus the number of LU factorizations at noise levels of (a) 10\%, (b) 20\%, and (c) 30\%. (d) Performance comparison of the final RME versus noise level.} 
%     \label{fig:BP_REG_MSE}
% \end{figure}
%
%
\paragraph{Colored noise scenario}
We generate colored noise by filtering AWGN through convolution with a Gaussian kernel.
Figure~\ref{fig:colored_noise} compares white (figure~\ref{fig:colored_noise}(a)) and colored (figure~\ref{fig:colored_noise}(b)) noise at the 30\% level. 
When the data is contaminated with colored noise, the assumption of AWGN underlying the RWP method is violated due to non-zero-lag correlation (figure~\ref{fig:colored_noise}(c)).
Under these conditions, the RGCV method with 7.76\% RME outperformed RWP with 8.22\% RME, as seen form figure~\ref{fig:colored_BP}, consistent with its demonstrated resilience under colored noise \citep{Bauer_2011_CCM}.
However, the generalized whiteness principle (GWP) \citep{Bevilacqua_2025_GFW} method can be adapted to improve the RWP. By applying techniques like downsampling to reduce non-zero-lag correlation bias (figure~\ref{fig:colored_noise}(c)), the accuracy of RWP is improved to 7.58\% RME (figure~\ref{fig:colored_BP}(d)).
In summary, we can see that RWP, when adapted using down-sampling, achieved the highest accuracy in the colored noise scenario.

 \subsection{Elastic FWI example} 

\begin{figure}  
  \centering  
  \includegraphics[width=.8\linewidth]{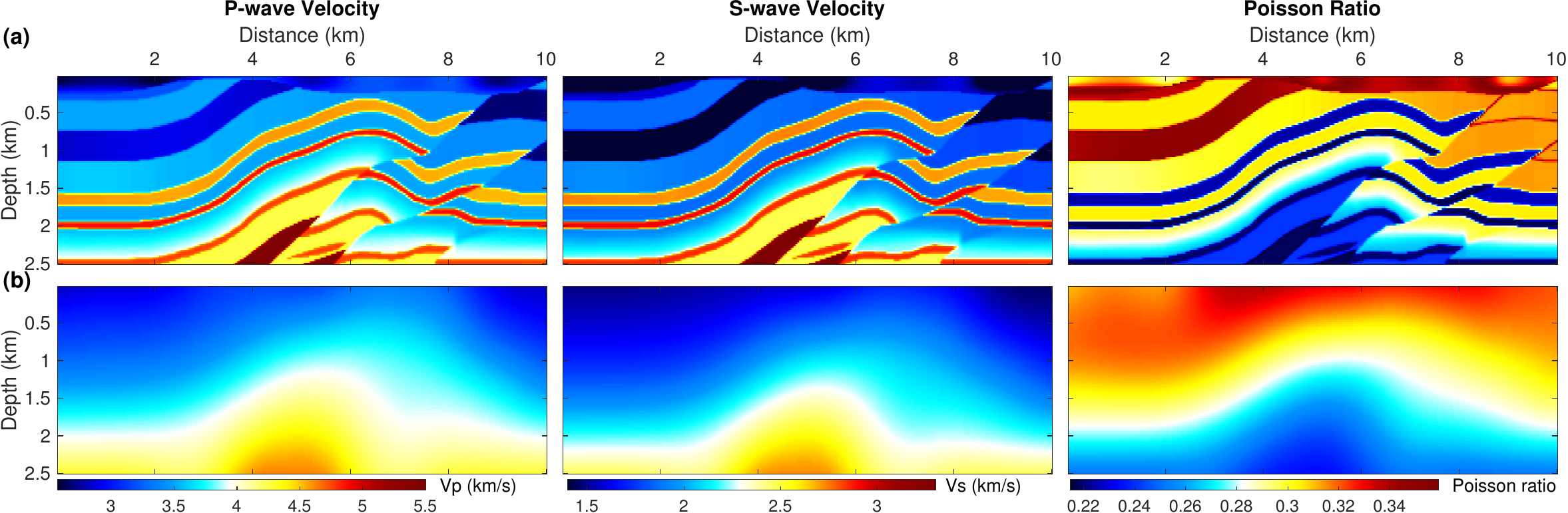} 
  \caption{Elastic FWI test. (a) True $\text{V}_\text{P}$, $\text{V}_\text{S}$, and Possion's ratio models. (b) Initial models used for inversion.}     \label{fig:El_models} 
  \end{figure}

 In this section, we evaluate the performance of the parameter selection strategies within the framework of elastic FWI, which involves simultaneously inverting for P-wave velocity ($V_P$) and S-wave velocity ($V_S$), and subsequently analyzing the resulting Poisson’s ratio model (figure~\ref{fig:El_models}(a)). The model complexity is high, featuring a spatially varying Poisson’s ratio.

\begin{figure}  
  \centering  
  \includegraphics[width=.6\linewidth]{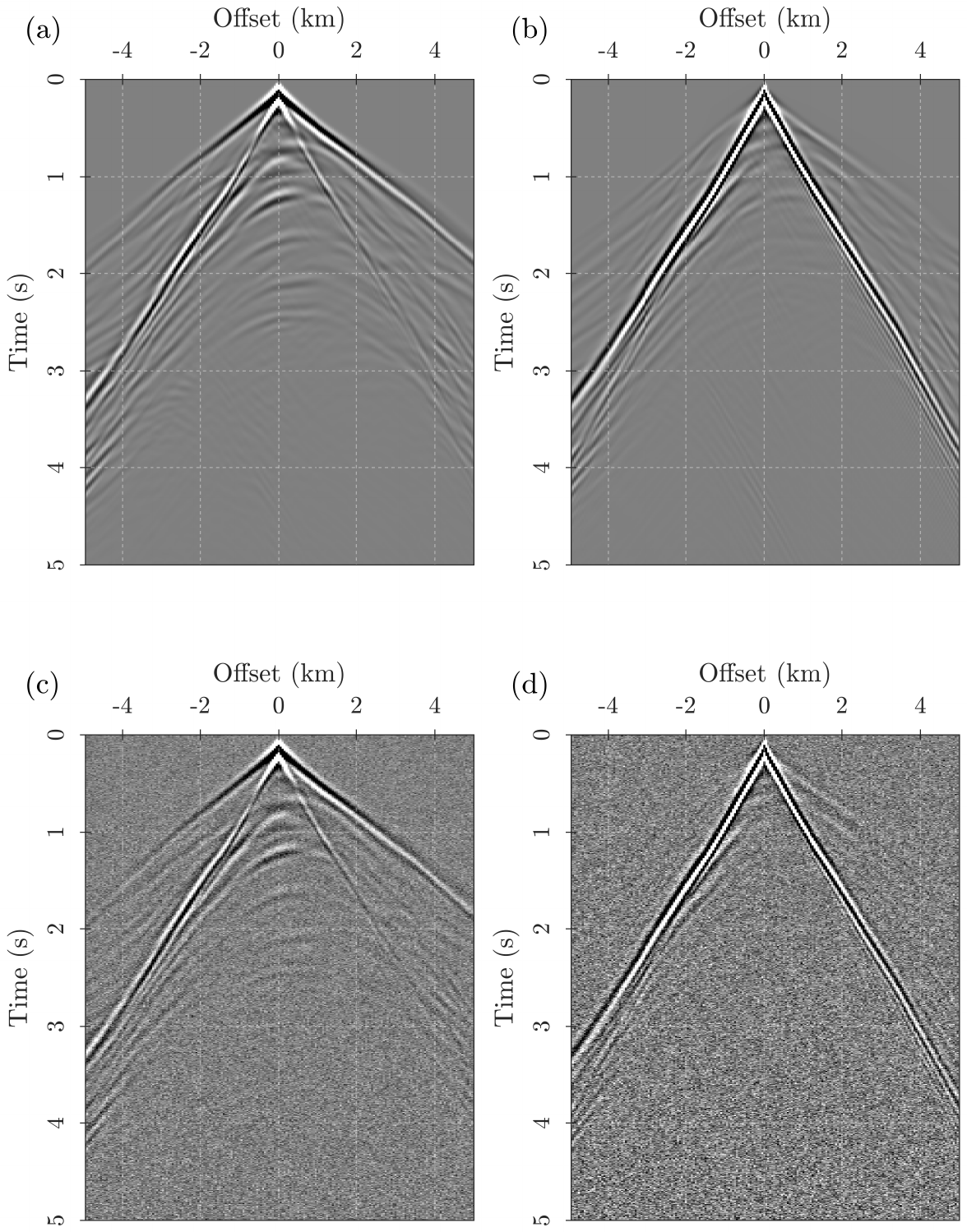} 
  \caption{Elastic FWI test. Time-domain comparison of the effect of additive Gaussian noise on a single shot gather, with the source located at the center of the model. (a--b) Noise-free reference data, $\bd_{x}$ and $\bd_{z}$, respectively. (c--d)  $\bd_{x}$ and $\bd_{z}$ contaminated with 30\% Gaussian noise (corresponding to SNR of 5~dB for  $\bd_{x}$ and 3~dB for  $\bd_{z}$), respectively.}
  \label{fig:El_data} 
  \end{figure}

We tested the inversion using data contaminated with $30\%$ complex Gaussian noise. Time-domain analysis (figure~\ref{fig:El_data}) confirms the severity of the noise, with the vertical component ($d_z$) reaching a low SNR of 3 dB. Frequency-domain inversion was performed over three cycles covering 3–13 Hz. We compare the Reduced, Penalty, and Dual-AL FWI formulations, each equipped with DP, RGCV, and RWP parameter selection methods, resulting in nine total configurations.

\subsubsection{Computational efficiency.}
The efficiency results consistently demonstrate the superiority of the Dual-AL formulation (Table~\ref{tab:EL_runtime}). We see that 1) The Reduced and Penalty approaches required 380 full PDE solves (requiring 380 LU factorizations across the 38 test frequencies). In contrast, the Dual-AL formulation utilized its multiplier-oriented framework to maintain a fixed background operator, requiring only 38 LU factorizations.
2) This massive reduction in the most expensive step translated directly into execution time savings. The Dual-AL method (runtime $\approx 3.1$ hours) was, on average, about 3.5 times faster than the Reduced and Penalty methods (runtime $\approx 10.6$ hours).

\subsubsection{Accuracy and RWP robustness.}
The Dual-AL formulation delivered the most accurate overall reconstruction, successfully recovering dipping layers and high-fidelity velocity contrasts (figures~\ref{fig:El_inv_res_DP}–\ref{fig:El_inv_res_RW}). Critically, its performance was highly dependent on the penalty parameter selection strategy:
1) For the Dual-AL method, RWP consistently outperformed both DP and RGCV, yielding the lowest overall reconstruction errors among all nine inversion schemes for both $V_P$ and $V_S$ (Table~\ref{tab:EL_runtime}). The final RME achieved by RWP was 12.48\% for $V_P$ and 16.37\% for $V_S$. The resulting Poisson’s ratio model (figure~\ref{fig:El_inv_res_RW}(c)) also exhibited the most well-resolved layering with reduced artifacts, confirming RWP's ability to maintain accuracy in multi-parameter problems.

\subsubsection{Parameter selection mechanism.}
\begin{figure} 
  \centering    
  \includegraphics[width=1\linewidth]{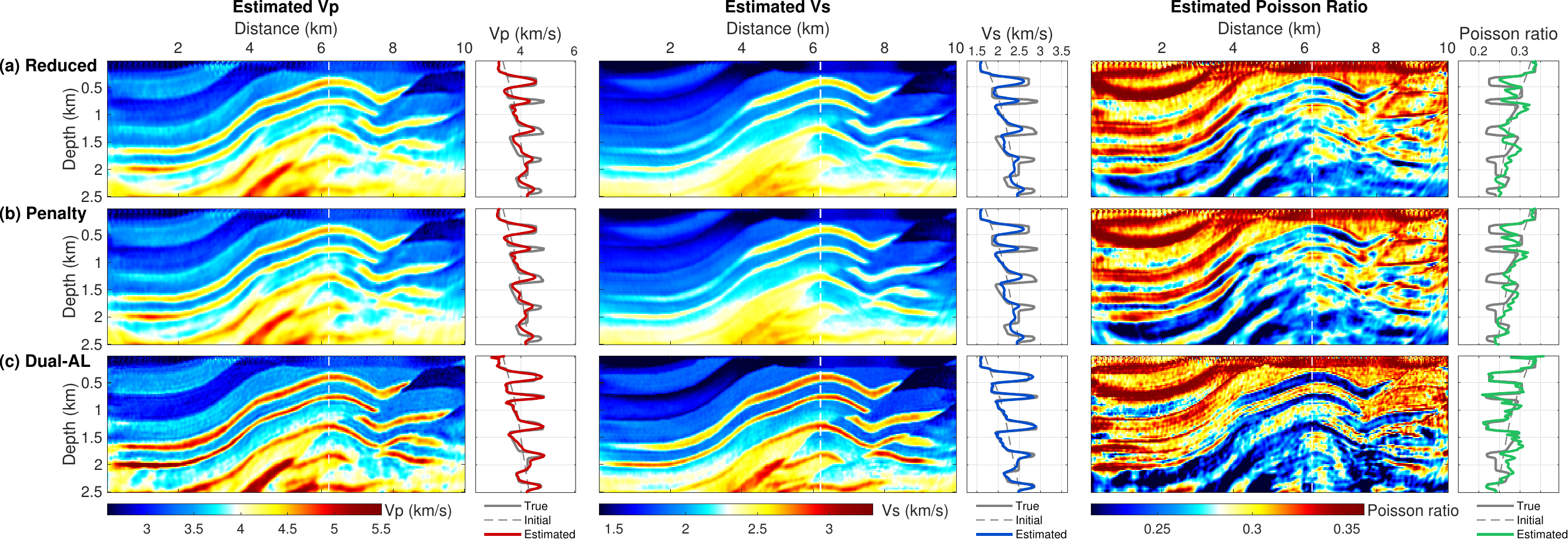}
  \caption{Elastic FWI test. Final inversion results obtained using the DP-based parameter selection strategy for three FWI formulations: 
    (a) Reduced, (b) Penalty, and (c) Dual-AL. Each row shows the estimated models for $\text{V}_\text{P}$ (left), $\text{V}_\text{S}$ (middle), and the computed Poisson’s ratio (right). Vertical profiles at 6.2 km compare the true (solid), initial (dashed), and inverted (colored) models.}  
  \label{fig:El_inv_res_DP} 
  \end{figure}

\begin{figure*} 
  \centering 
  \includegraphics[width=0.5\linewidth]{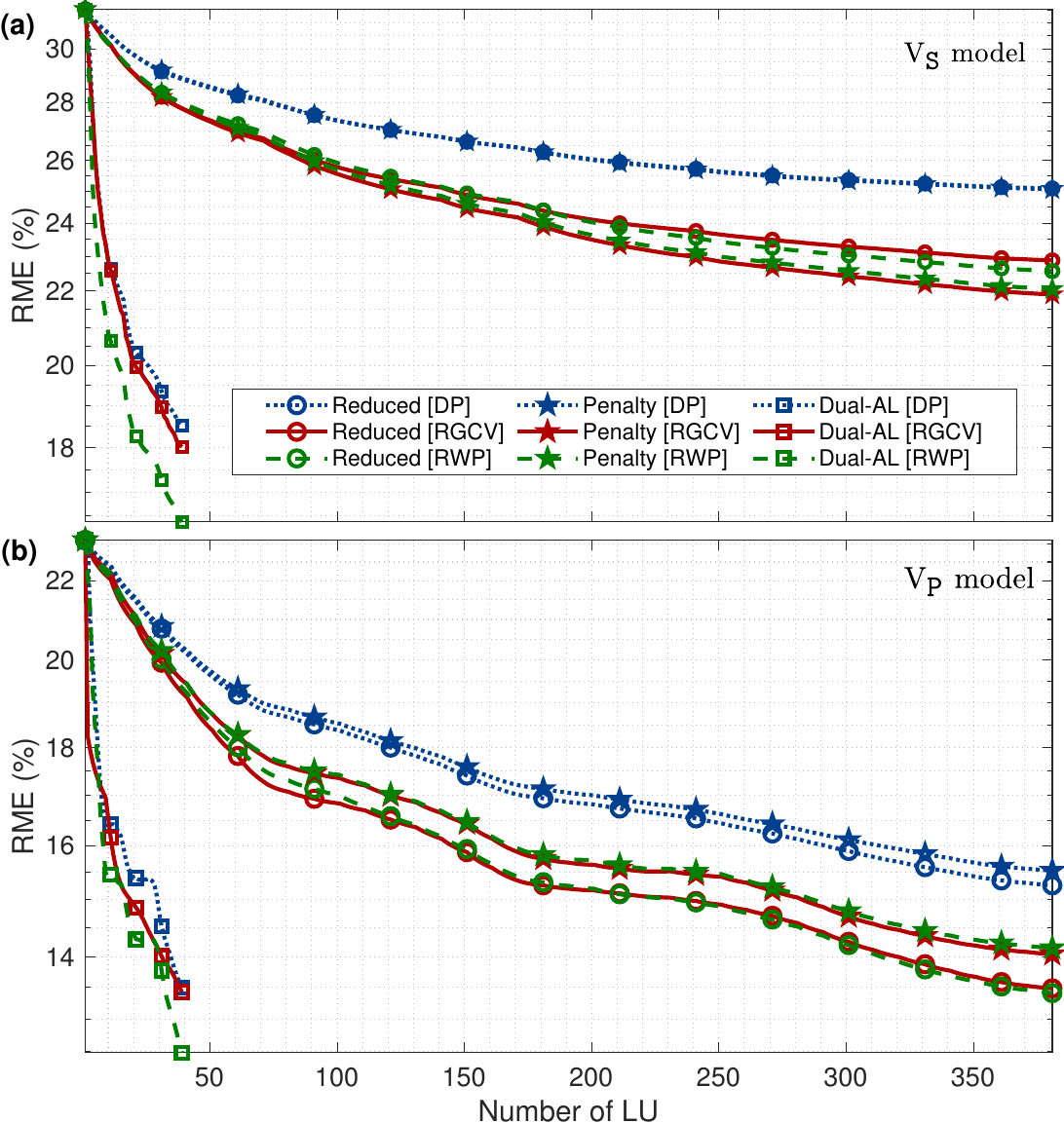}    
  \caption{The evolution of RME for (a) S-wave velocity, and (b) P-wave velocity for different elastic FWI methods using DP, RGCV, and RWP as parameter selection tools.}
  \label{fig:MSE_elastic}
  \end{figure*}

\begin{figure*}
  \centering    
  \includegraphics[scale=0.6]{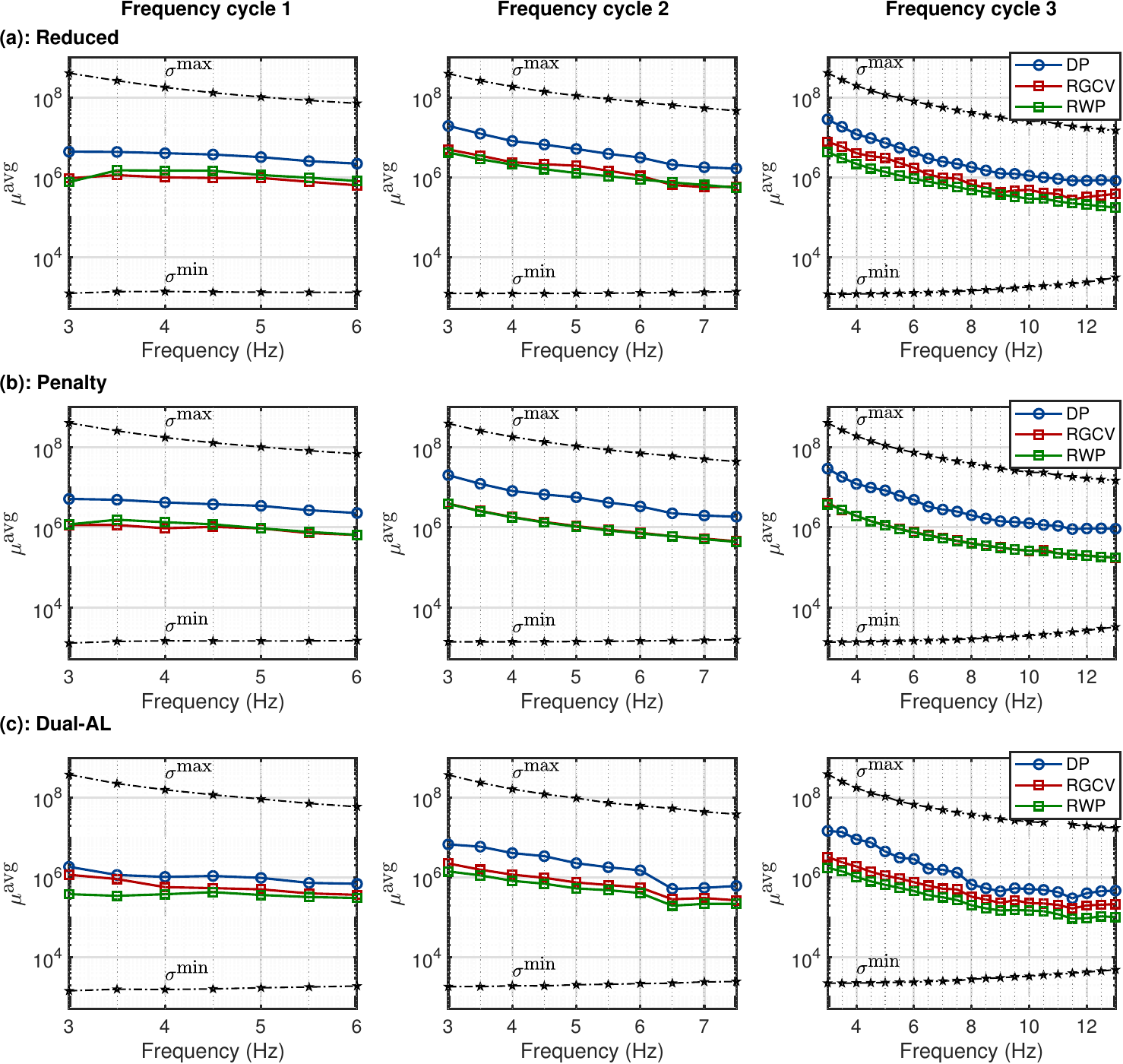}    
  \caption{Elastic FWI test. Average regularization parameter ($\mu^{\text{avg}}$) computed using three parameter selection methods—DP, RGCV, and RWP. The values are averaged over the inner iterations associated with the update of $\varepsilon$ for each test frequency and noise level. Results are shown for (a) Reduced, (b) Penalty, and (c) Dual-AL methods across three frequency cycles. The minimum and maximum singular values of $\bS\bS^{\top}$ ($\sigma^{\text{min}}$, $\sigma^{\text{max}}$) are also included.} 
  \label{fig:mu_vals_elastic}
  \end{figure*}

This superior accuracy is explained by RWP’s regularization philosophy. Figure \ref{fig:mu_vals_elastic} confirms that DP consistently selects larger $\mu$ values than the other methods, leading to overly smooth results (figure~\ref{fig:El_inv_res_DP}). For the high-performing Dual-AL method (figure~\ref{fig:mu_vals_elastic}(c)), RWP consistently produces slightly smaller parameters than RGCV. This subtle difference in penalty strength, enabled by RWP’s whiteness criteria, allows Dual-AL to achieve more accurate reconstructions, as verified by the RME analysis (figure~\ref{fig:MSE_elastic}).

\begin{figure} 
  \centering  
  \includegraphics[width=1\linewidth]{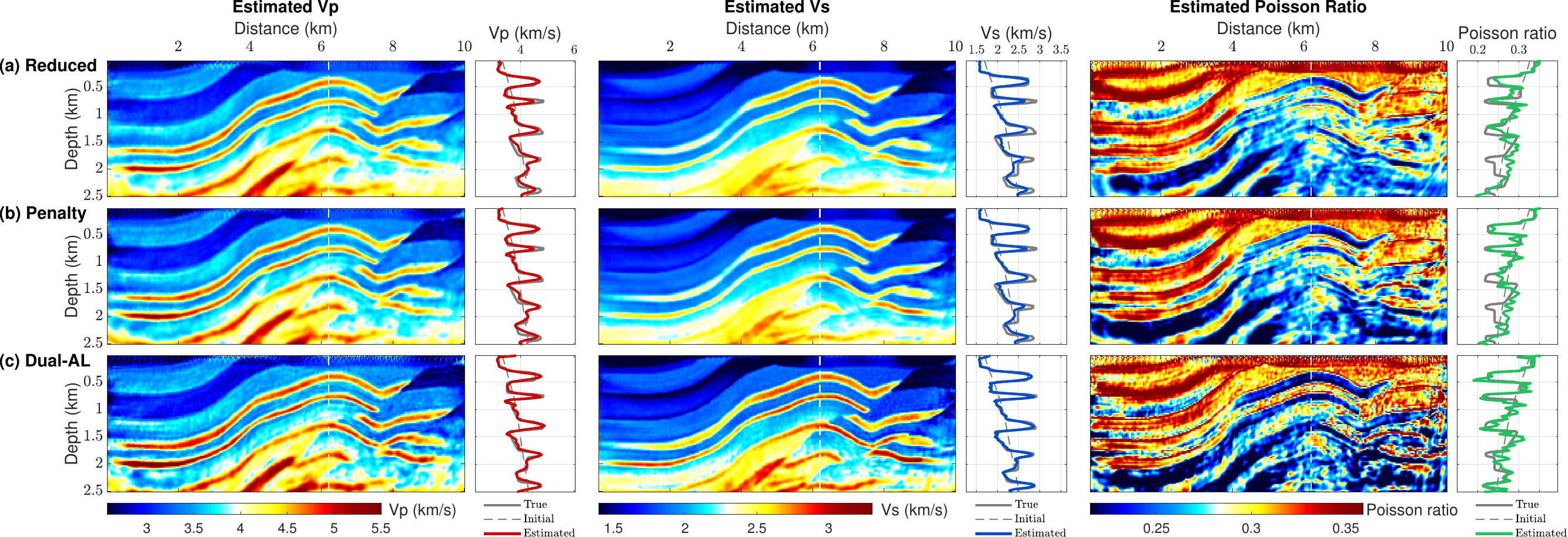}   
  \caption{Elastic FWI test. Same as figure~\ref{fig:El_inv_res_DP} employing RGCV strategy.}  
  \label{fig:El_inv_res_GCV}
  \end{figure}

\begin{figure}  
  \centering     
  \includegraphics[width=1\linewidth]{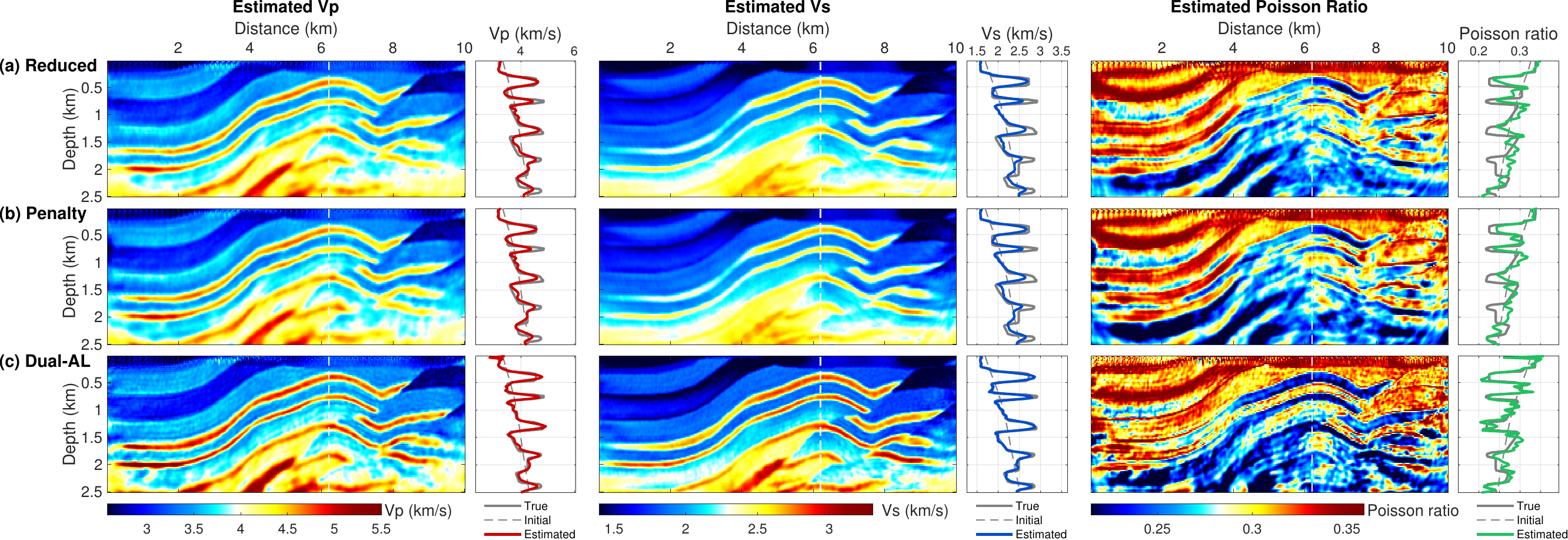}   
  \caption{Elastic FWI test. Same as figure~\ref{fig:El_inv_res_DP} employing RWP strategy.}  
  \label{fig:El_inv_res_RW}
  \end{figure}

\begin{figure}  
 \centering   
\includegraphics[scale=0.35]{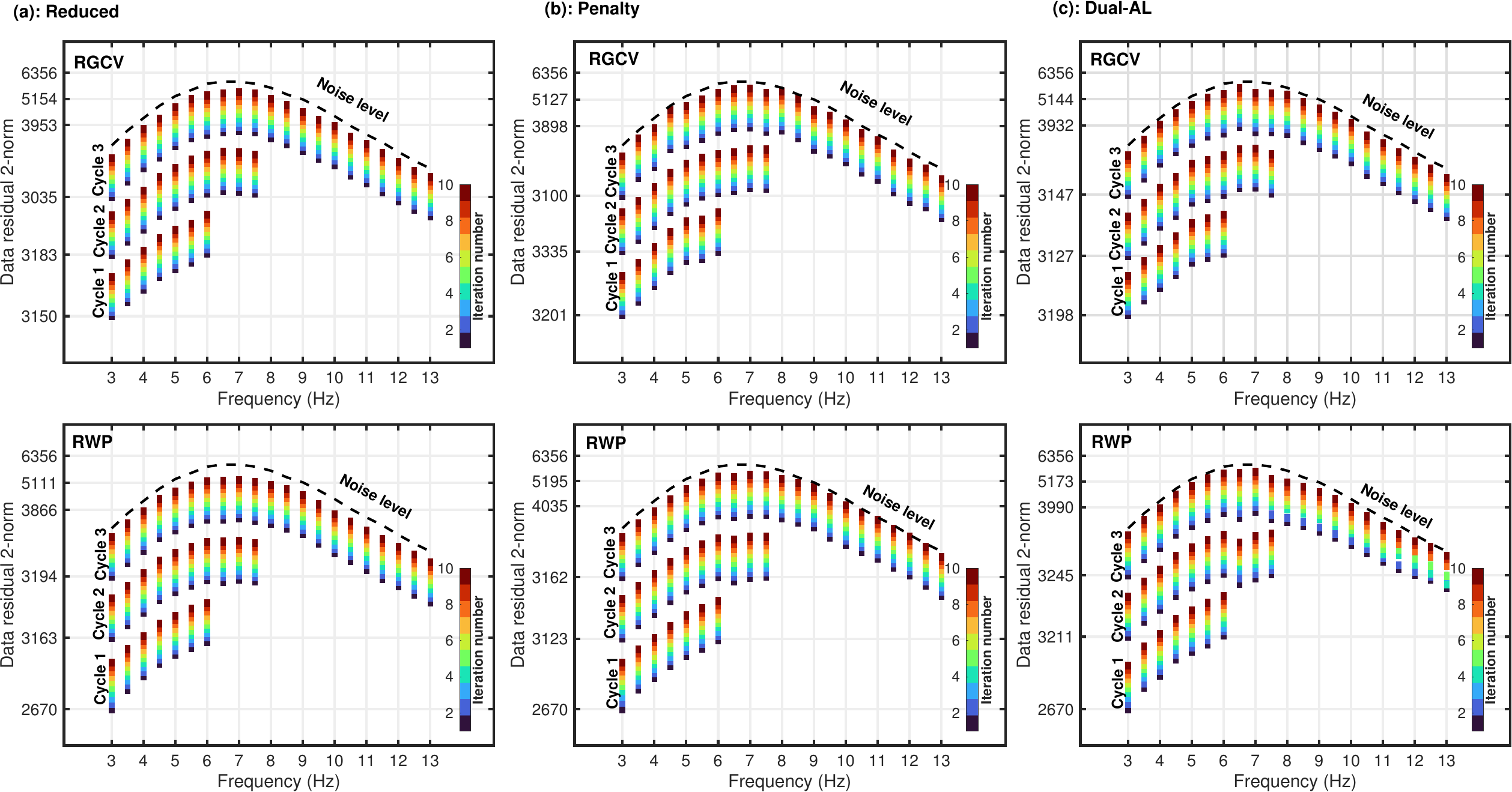}   
 \caption{Elastic FWI test. Evolution of 2-norm data residuals during iterations for different FWI strategies across frequencies. The plots display data residual values versus frequency for three FWI appproaches: (a) Reduced-space method, (b) Penalty method, and (c) Dual-AL method. Results are shown for two regularization parameter selection strategies: RGCV  (top row) and RWP (bottom row). Colors represent different iteration numbers during the inversion process, progressing from early iterations (cooler colors) to later iterations (warmer colors) as indicated by the color bar (iteration number 1-10). The dashed curves indicate noise levels at each frequency.}   
\label{fig:El_misfit}
\end{figure}

The reduced and penalty formulations, due to their inherent computational intensity, yielded inferior results regardless of the parameter selector used, producing blurred interfaces and poorly resolved boundaries (figures~\ref{fig:El_inv_res_DP}(a), \ref{fig:El_inv_res_GCV}(a), \ref{fig:El_inv_res_RW}(a)). The convergence of the data residual (figure~\ref{fig:El_misfit}) further confirms Dual-AL's superior performance in fitting the noisy data across all frequencies compared to the Reduced and Penalty methods.


  %

%To compare the regularization parameters selected by each method, Fig.~\ref{fig:mu_vals_elastic} shows the average parameter value estimated for each frequency inversion. For reference, the minimum ($\sigma^{\text{min}}$) and maximum ($\sigma^{\text{max}}$) eigenvalues of the data-space Hessian are also plotted. We see that
%i) In all cases, the selected parameters lie within the range $[\sigma^{\text{min}}, \sigma^{\text{max}}]$.
%%iii) For the penalty method (Fig.~\ref{fig:mu_vals_elastic}b), GCV and RWP yield nearly identical $\mu$ values across all frequency cycles, resulting in comparable inversion accuracy, as confirmed by the relative model error curves in Fig.~\ref{fig:MSE_elastic}.
%iv) For the dual-AL method (Fig.~\ref{fig:mu_vals_elastic}c), DP and GCV behave more similarly than in the reduced and penalty cases, although DP still trends toward larger $\mu$ values. In contrast, RWP consistently produces slightly smaller parameters than GCV. This subtle difference enables dual-AL to achieve more accurate reconstructions, as reflected in both the Poisson’s ratio maps and the error analysis (Figs.~\ref{fig:El_inv_res_RW}–\ref{fig:MSE_elastic}).

%
%
%

  % \begin{table} % \caption{Performance of different FWI methods using DP, GCV, and RWP-based parameter selection strategies for the elastic example.} % \label{tab:BP_runtime} % \centering   % \begin{tabular}{l c c c c c c c c c}  % \toprule  % & \multicolumn{3}{c}{Reduced} & \multicolumn{3}{c}{Penalty} & \multicolumn{3}{c}{Dual-AL} \\  % \cmidrule(lr){2-4} \cmidrule(lr){5-7} \cmidrule(lr){8-10} % & LU.no & RMSE & Runtime & LU.no & RMSE & Runtime & LU.no & RMSE & Runtime \\  % \midrule  % Parameter selection & & & & & & & & & \\ % \quad DP  & 380 & 00 & 10.55 & 380 & 00 & 10.64 & 38 & 00 & 3.04 \\ % \quad GCV & 380 & 00 & 10.63 & 380 & 00 & 10.68 & 38 & 00 & 3.09 \\ % \quad RWP & 380 & 00 & 10.63 & 380 & 00 & 10.69 & 38 & 00 & 3.12 \\ % \bottomrule  % \end{tabular} %\ end{table}  
  
\begin{table}
    \caption{Performance of different elastic FWI methods using DP, RGCV, and RWP-based 
    parameter selection strategies.}
    \label{tab:EL_runtime}
    \centering
    \begin{adjustbox}{width=\textwidth}
    \begin{tabular}{l c c c c c c c c c c c c c}
        \toprule
        & \multicolumn{4}{c}{Reduced} 
        & \multicolumn{4}{c}{Penalty} 
        & \multicolumn{4}{c}{Dual-AL} \\
        \cmidrule(lr){2-5} \cmidrule(lr){6-9} \cmidrule(lr){10-13}
        & \multirow{2}{*}{LU No.} 
        & \multirow{2}{*}{Runtime (h)} 
        & \multicolumn{2}{c}{RME (\%)} 
        & \multirow{2}{*}{LU No.} 
        & \multirow{2}{*}{Runtime (h)} 
        & \multicolumn{2}{c}{RME (\%)} 
        & \multirow{2}{*}{LU No.} 
        & \multirow{2}{*}{Runtime (h)} 
        & \multicolumn{2}{c}{RME (\%)} \\
        \cmidrule(lr){4-5} \cmidrule(lr){8-9} \cmidrule(lr){12-13}
        & & & ($\text{V}_\text{P}$) & ($\text{V}_\text{S}$) 
          & & & ($\text{V}_\text{P}$) & ($\text{V}_\text{S}$) 
          & & & ($\text{V}_\text{P}$) & ($\text{V}_\text{S}$) \\
        \midrule
        Parameter selection & & & & & & & & & & & & \\
        \quad DP  
            & 380 & 10.55 & 15.26 & 25.08 
            & 380 & 10.64 & 15.53 & 25.06 
            & 38  & 3.04  & 13.49 & 18.52 \\
        \quad RGCV 
            & 380 & 10.63 & 13.48 & 22.88 
            & 380 & 10.68 & 14.05 & 21.90 
            & 38  & 3.09  & 13.42  & 18.02 \\
        \quad RWP 
            & 380 & 10.63 & 13.40 & 22.57 
            & 380 & 10.69 & 14.15 & 22.05 
            & 38  & 3.12  & 12.48 & 16.37 \\
        \bottomrule
    \end{tabular}
    \end{adjustbox}
\end{table}

%\bibliographystyle{gji.bst}
%\bibliography{WINTER_BIBLIO_TEMP,WINTER_BIBLIO}

\section{Conclusions}
The integration of the Residual Whiteness Principle (RWP) into the multiplier-based FWI formulation successfully addresses the critical challenge of penalty parameter ($\mu$) selection, which significantly impacts solution convergence and robustness in noisy environments. Unlike the traditional Discrepancy Principle (DP), which requires an accurate and often unknown estimate of the noise variance ($\sigma$) and considers only the zero-lag residual autocorrelation, RWP eliminates the reliance on $\sigma$ by evaluating the full autocorrelation function to enforce whiteness in the predicted residual. This dynamic selection strategy was embedded within the highly efficient Dual-AL method. This dual-space framework is essential because it allows the background wave equation operator to remain fixed, requiring only a single LU matrix factorization per frequency inversion. This design enabled the dynamic adjustment of $\mu$ within each iteration at negligible computational cost, rendering the resulting algorithm scalable and practical for large-scale applications. Numerical experiments consistently confirmed the computational advantage, with the Dual-AL formulation achieving approximately a 3.5 times speedup over traditional reduced and penalty methods.

The empirical results across complex acoustic and elastic benchmark models demonstrate the superior scientific strength and robustness of the RWP strategy when dealing with noise. RWP exhibited exceptional noise robustness under Additive White Gaussian Noise (AWGN), maintaining a nearly constant reconstruction error across all tested noise levels, and achieved the closest approximation to the optimal parameter choice among the methods. Conversely, DP degraded significantly, despite being provided the exact noise variance, due to its selection of overly large $\mu$ values that lead to excessive smoothing. Furthermore, in complex multi-parameter elastic FWI, RWP consistently produced the lowest reconstruction errors for both P- and S-wave velocities. These findings establish RWP as a reliable, efficient, and fully automatic framework for wave-based inversion, offering a practical solution, particularly where measurement noise is significant and its characteristics are unknown

% \section{Acknowledgments}  
% This research was financially supported by the SONATA BIS grant
% (No. 2022/46/E/ST10/00266) of the National Science Center in
% Poland. 
%
% Each of the commands below will create an unnumbered section with the appropriate heading.
% Remove any sections that are not relevant for your article.
% All sections except suppdata will be removed if the [anonymous] option is used.
% See iopjournal-guidelines.pdf for more information.
%

% \ack{Sample text inserted for demonstration.}
\begin{acknowledgments}
This research was financially supported by the SONATA BIS grant
(No. 2022/46/E/ST10/00266) of the National Science Center in
Poland.
\end{acknowledgments}

\begin{dataavailability}
The data that support this study are available from the corresponding
author on reasonable request.
\end{dataavailability}

% This section is a list of funder names and grant numbers

% \roles{Sample text inserted for demonstration.}
% List author names and the contributions made to the article, using terms from the NISO Contributor Roles Taxonomy (CRediT) https://credit.niso.org

% \data{Sample text inserted for demonstration.}
% For more information on IOP Publishing's research data policy see: https://publishingsupport.iopscience.iop.org/questions/research-data/

% \suppdata{Data supporting this study’s findings are available from the authors on reasonable request.}

\bibliographystyle{gji.bst} % Vancouver bibliography style
%\bibliography{WINTER_BIBLIO_TEMP,WINTER_BIBLIO}

\newcommand{\SortNoop}[1]{}

%\section*{References}

\end{document}